\newif\ifTWO
\global\TWOtrue

\ifTWO
\documentclass[preprint2]{aastex}
\else
\documentclass[manuscript]{aastex}
\fi
\usepackage{epsfig}
\usepackage{url}

\usepackage{color}    
\newcommand{\tcx}{\textcolor{black}}

\newcommand{\D}{\Delta}
\newcommand{\Dt}{\mbox{$\D t$}}
\newcommand{\dt}{\mbox{$\delta t$}}
\newcommand{\eM}{\mbox{$e_{\cal M}$}}
\newcommand{\DeM}{\mbox{$\Delta e_{\cal M}$}}
\newcommand{\rM}{\mbox{$|\v{r}_{\cal M}|$}}
\newcommand{\drM}{\mbox{$|\delta \v{r}_{\cal M}|$}}
\newcommand{\eMx}{\mbox{$e_{\cal M}^{\rm max}$}}
\newcommand{\up}{\uparrow}
\newcommand{\aM}{\mbox{$a_{\cal M}$}}
\newcommand{\ms}{{\tt mercury6}}
\newcommand{\hnb}{{\tt HNBody}}
\newcommand{\Sms}[1]{\mbox{$S^{\rm m6}_{#1}$}}
\newcommand{\Shnb}[1]{\mbox{$S^{_{\rm HNB}}_{#1}$}}
\newcommand{\Q}{\cal Q}
\renewcommand{\P}{\cal P}
\newcommand{\dv}{\mbox{$\delta{\v{v}}$}}
\newcommand{\qt}{\mbox{\small$\q{1}{2}$}}
\newcommand{\e}[1]{\mbox{$\x10^{#1}$}}
\newcommand{\x}{\times}
\newcommand{\bm}{\boldmath}
\newcommand{\ubm}{\unboldmath}
\renewcommand{\v}[1]{\mbox{\bm$#1$\ubm}}
\newcommand{\beqn}{\begin{eqnarray}}
\newcommand{\eeqn}{\end{eqnarray}}
\newcommand{\beq}{\begin{eqnarray*}}
\newcommand{\eeq}{\end{eqnarray*}}
\newcommand{\scs}{\scriptsize}
\newcommand{\pmo}{\mbox{$^{-1}$}}
\newcommand{\q}{\frac}
\newcommand{\lrarr}{\longrightarrow}
\newcommand{\nn}{\nonumber}
\newcommand{\prt}{\partial}

\shorttitle{Solar System Dynamic Stability}
\shortauthors{Zeebe}

\begin{document}


\title{
Dynamic stability of the Solar System: Statistically 
inconclusive results from ensemble integrations 
}

\author{Richard E. Zeebe$^{1,*}$}
\affil{\vspace*{0.5cm}
     $^*$Corresponding Author.\\ 
     $^1$School of Ocean and Earth Science and Technology, 
     University of Hawaii at Manoa, 
     1000 Pope Road, MSB 629, Honolulu, HI 96822, USA. 
     email: zeebe@soest.hawaii.edu\\[4ex]
     {\bf The Astrophysical Journal}, accepted, Oct 08, 2014
     }

\begin{abstract}
Due to the chaotic nature of the Solar System, the question 
of its long-term stability can only be answered in a statistical 
sense, for instance, based on numerical ensemble integrations of 
nearby orbits. Destabilization of the inner planets, leading to 
close encounters and/or collisions can be initiated through a 
large increase in Mercury's eccentricity, with a currently 
assumed likelihood of $\sim$1\%. However, little is known at 
present about the robustness of this number.
Here I report ensemble integrations of the full equations 
of motion of the eight planets and Pluto over 5~Gyr, 
including contributions from general relativity.
The results show that different numerical algorithms lead 
to statistically different results for the evolution of 
Mercury's eccentricity (\eM). For instance, starting 
at present initial conditions ($\eM \simeq 0.21$), Mercury's 
maximum eccentricity achieved over 5~Gyr is on average significantly 
higher in symplectic ensemble integrations using heliocentric
than Jacobi coordinates and stricter error control. In contrast, 
starting at a possible future configuration ($\eM \simeq 0.53$), 
Mercury's maximum eccentricity achieved over the subsequent 
500~Myr is on average significantly lower using heliocentric than 
Jacobi coordinates. For example, the probability
for \eM\ to increase beyond 0.53 over 500~Myr is 
$>$90\% (Jacobi) vs.\ only 40-55\% (heliocentric). This poses 
a dilemma as the physical evolution of the real system ---
and its probabilistic behavior --- cannot depend on 
the coordinate system or numerical algorithm chosen to 
describe it. Some tests of the numerical algorithms
suggest that symplectic integrators using heliocentric
coordinates underestimate the odds for destabilization 
of Mercury's orbit at high initial \eM.
\end{abstract}


\newpage

\section{Introduction}

The question whether the Solar System is dynamically stable 
over long periods of time has received considerable 
attention for centuries, 
including contributions from Newton, Lagrange, Laplace, 
Poincar{\'e}, Kolmogorov, Arnold, Moser etc.\ 
\citep[e.g.][]{laskar13}. Research in this field has recently 
experienced a renaissance due to advances in
computational power and numerical algorithms,
permitting integration of the full equations of motion
approaching the Solar System's lifetime ($\pm\sim$5~Gyr)
\citep{wisdom91,quinn91,sussman92,saha92,murray99,
ito02,varadi03,batygin08,laskar09}.
Moreover, parallel computing now allows tackling
the problem of stability with statistical means through 
simultaneous
integration of multiple nearby orbits \citep{laskar09}.
A statistical approach is necessary because of the
chaotic behavior of the system, i.e.\ the sensitivity 
of orbital solutions to initial conditions \citep{laskar89,
sussman92,murray99,richter01,varadi03,batygin08}. Small differences
in initial conditions grow exponentially, with a 
time constant (Lyapunov time)
for the inner planets of only $\sim$5~Myr 
\citep{laskar89,varadi03,batygin08}. 
\tcx{For example, a difference in initial coordinates
of 1~mm grows to $\sim$1~AU (=~1.496\e{11}~m) 
after 163~Myr.} Thus, the chaotic nature of the planetary 
orbits makes it fundamentally impossible to predict their 
evolution accurately beyond $\sim$100~Myr 
\citep{laskar89}. Hence the stability 
problem can only be answered in probabilistic terms,
e.g.\ by studying the behavior of a large number of 
physically possible solutions.
The quest for a single deterministic solution, which
conclusively describes the Solar System's evolution
for all times \citep[in the spirit of Laplace's demon,]
[]{laplace51}, must be regarded as quixotic.

An ensemble integration of 2,501 nearby orbits over 5~Gyr 
has recently been reported based on the full 
equations of motion and including contributions from 
general relativity and the Moon \citep{laskar09}.
In those simulations, two adjacent orbits differed
initially by only 0.38~mm in Mercury's semi-major axis 
(\aM). The largest overall offset in initial \aM\ was 
hence $\rm2,500\x0.38~mm=0.95$~m, 
well within the uncertainty of 
our current knowledge of the Solar System. In about
1\% of the simulations, a large increase in Mercury's 
eccentricity was reported, which can lead to destabilization 
of the inner planets, including close encounters 
and/or collisions. Note that the obtained probability 
distribution of Mercury's eccentricity was similar to 
results obtained earlier with averaged equations 
\citep{laskar08}.
Most surprisingly, further simulations 
included the possibility of a collision between Earth 
and Venus via transfer of angular momentum from the giant 
planets to the terrestrial planets \citep{laskar09}.
Given the chaotic behavior of the system, such an 
outcome might be within the range of possibilities.
However, little is currently known about the potential
dependence of the simulated trajectories and statistical 
results on the numerical integrator, step size, 
integration coordinates, etc.

Here I report ensemble integrations of Solar System 
orbits over 5~Gyr, including contributions from 
general relativity. The simulations reveal a strong 
influence of integration coordinates on statistical 
results, i.e.\ on the predicted odds for destabilization 
of Mercury's orbit. Furthermore, I discuss resonances
and Lyapunov times, and perform several 
tests aiming at resolving the dilemma of the dependency 
of statistical results on integration coordinates.

\section{Methods}

The full equations of motion of the eight planets and Pluto 
were integrated over 5~Gyr into the future using 
the numerical integrator packages \ms\ and \hnb\ and 
various integration options \citep{chambers99,rauch02}.
\tcx{Unless stated otherwise, symplectic integrators 
were used. Symplectic integrators exactly describe the time 
evolution of a (modified) Hamiltonian system that is very close 
to the true Hamiltonian \citep[cf.\ e.g.][]{wisdom91,yoshida93,
chambers99} and hence do not suffer from long-term
buildup in energy error. The maximum energy 
variation along a given orbit in symplectic integrations 
(see below) provides a measure of the difference between 
the modified and the true Hamiltonian.}
Relativistic corrections are critical \citep{laskar09}
and were available in \hnb\
but not in \ms. Post-Newtonian corrections for 
symplectic integration \citep{mikkola98,soffel89} were therefore 
implemented before using \ms\ (see Appendix). Thus, all simulations 
presented here include contributions from general relativity,
unless explicitly stated. To allow comparison with previous 
studies, several higher-order effects were not included 
here. This means ignoring potential effects from asteroids 
\citep{ito02,batygin08}, perturbations from passing stars, 
and solar mass loss \citep{ito02,batygin08,laskar09}.
Furthermore, the Earth-Moon system was considered
a single mass point, located at the Earth-Moon barycenter.

Per ensemble integration, 40 orbital solutions were 
computed with each package, starting from the same set 
of initial conditions, where Mercury's initial 
radial distance was offset by 7~cm between every two 
adjacent orbits (largest overall offset: 
$\rm39\x7~cm=2.73$~m). Initial conditions for all bodies
in the 5-Gyr runs (before offsetting Mercury)
were generated from DE431
(\url{naif.jpl.nasa.gov/pub/naif/generic_kernels/spk/planets})
at JD2451544.5 (01 Jan 2000) using the SPICE toolkit for Matlab 
(\url{naif.jpl.nasa.gov/naif/toolkit.html}) 
(Table~\ref{TabDE431}).

For the 5-Gyr simulations with \ms, the hybrid integrator 
with democratic-heliocentric coordinates 
\citep{chambers99,duncan98}
following \citet{batygin08} and a 6-day initial 
timestep (\Dt) for the 2nd order symplectic algorithm was
used. In case Mercury's eccentricity (\eM) increased above certain 
threshold values during the 5-Gyr simulations with \ms,
\Dt\ was reduced but held constant after that until the 
next threshold was crossed (if applicable). 
For example, starting at $\D t = 6$~days
and $\eM = 0.21$, \Dt\ was reduced to 5~days when \eM\
exceeded 0.34. \ms's source code is available 
\citep{chambers99} and was manipulated accordingly. 
The threshold values and corresponding timesteps 
(Table~\ref{TabEcrit}) were chosen to maintain a roughly 
similar maximum relative energy error (${\rm max} 
|\D E/E| = {\rm max} |(E(t)-E_0)/E_0|$) throughout the 
simulation. \hnb's 
source code is not available \citep{rauch02} and a constant
4-day timestep was used throughout the 5-Gyr simulations, 
employing a 2nd order symplectic integrator with corrector
and Jacobi coordinates \citep[see below and][]{wisdom91}.

\begin{table}[hhhhhh]
\begin{center}
\caption{Threshold \eM\ values and timesteps for the 
5-Gyr simulations with \ms. \label{TabEcrit}}
\begin{tabular}{lrrrrrr}
\tableline\tableline
$\eM >$    & 0.34 & 0.38 & 0.44 & 0.49 & 0.57 & 0.62 \\
\hline     
$\D t$ (d) & 5    & 4    & 3    & 2    & 1.5  & 1    \\
\tableline
\end{tabular}
\end{center}
\end{table}

The 40 orbital solutions were simultaneously integrated on a
64-bit Linux cluster (one 5-Gyr job per core on
10 Intel i7-3770 \@3.40~GHz nodes with 4~cores each).
Typical integration times for the 5-Gyr runs with \ms\ were 
$12-23$~days wall-clock time depending on step-size
reduction ($\Dt \leq 6$~d, see above) and 12~days wall-clock 
time with \hnb\ ($\Dt = 4$~d). 
Integrations with \hnb\ were performed using the pre-compiled 
binary version for Linux 64-bit Intel/AMD64. Executables
of \ms\ were compiled from source code on 64-bit Linux platforms 
using {\tt gfortran 4.6.3}. When compared to test runs with \ms\ 
on a true 32-bit machine, a small long-term decrease in angular
momentum was noticeable in the output of the 64-bit \ms\
executable at small step size. The issue disappeared when 
using one of the following options on 64-bit platforms. 
(1) Compile with flag {\tt -m32}
(generates code for a 32-bit environment but is slower).
(2) In the \ms\ subroutine \verb|drift_kepmd()| replace
the original approximations for $\sin(x)$ and $\cos(x)$
by actual $\sin(x)$ and $\cos(x)$ functions.

\subsection{Democratic-heliocentric- and Jacobi coordinates}

The effect of different integration coordinates on the 
results of the symplectic integrations will be discussed 
at length below. Hence the definition of the coordinates
is given here. Democratic-heliocentric coordinates simply 
consist of heliocentric (bodycentric) positions and 
barycentric momenta \citep{duncan98,chambers99},
see Section~\ref{SecTwoBdy}.

Jacobi coordinates have a slightly more complex structure 
but are useful in symplectic integrations of the $n$-body 
problem because they allow writing the kinetic energy as a
diagonal sum of squares of the momenta \citep{wisdom91}.
In Jacobi coordinates, the first coordinate, $\v{Q}_0$, 
may be taken as the position of the center of mass. 
The first relative coordinate is the position of the first 
planet relative to the central body. The second 
relative coordinate is the position of the second
planet relative to the center of mass of the central
body and the first planet, and so forth. Thus, the 
$i$th relative coordinate is the position of the $i$th
planet relative to the center of mass of the central
body and the planets with lower indices:
\beqn
\v{Q}_i = \v{x}_i - \v{X}_{i-1} \ ,
\eeqn
where $0 < i < n$ and:
\beqn
\v{X}_{i}  = M_i^{-1} \ \sum_{j=0}^i m_j \v{x}_j
\qquad ; \qquad
M_i        = \sum_{j=0}^i m_j
\eeqn
is the center of mass of bodies with indices up to $i$.
The momenta are:
\beqn
\v{P}_i = m'_i \v{V}_i \ ,
\eeqn
where $\v{V}_i$ is the time derivative of $\v{Q}_i$
and $m'_i = M_{i-1} m_i / M_{i}$ for $0 < i < n$ and 
$m'_0 = M_{n-1} = M$, is the sum of all masses.
Interestingly, because of the Sun's dominant mass, the
difference in the actual values of heliocentric and 
Jacobi coordinates is small in the Solar System (e.g.\ zero 
for Mercury and of order $10^{-7}$ and $10^{-6}$~AU for 
Venus and Earth). Yet the difference in the results
of symplectic integrations due to these different 
sets of coordinates is significant (see below).
\tcx{Note that the comment above is meant for illustrative
purposes only. All initial input coordinates for 
\ms\ and \hnb\ were specified in heliocentric 
positions and velocities (e.g.\ Table~\ref{TabDE431}). 
Transformation to the desired
set of integration coordinates is performed internally.
In the following, "same initial conditions" refers 
to truly identical initial conditions, regardless of the 
integration coordinates used.}

\section{Results: 5-Gyr and 500-Myr simulations}

At first glance, the 5-Gyr simulations with \ms\ and 
\hnb\ appeared to yield similar 
results, except for different energy and momentum errors
(Fig.~\ref{FigAll5gyr}). In none of the 80 simulations
a large increase in \eM\ or a destabilization of the 
inner solar system occurred, which is also not expected
if the corresponding odds are $\sim$1\% \citep{laskar09}.
However, at second glance differences between the
\ms\ and \hnb\ runs became apparent.
For runs with identical initial conditions, the maximum 
difference in Mercury's eccentricity (\DeM) between 
the \ms\ and \hnb\ simulations typically
grew from $\sim$$10^{-7}$ to 0.01 over only $\sim$10~Myr.
For comparison, when \hnb\ was run with bodycentric 
coordinates (all else equal), \DeM\ grew to 0.01 over 
$\sim$60~Myr (see below).
Furthermore, \DeM\ between two simulations with the same 
package and options but different initial conditions
grew to 0.01 over $\sim$75~Myr and $\sim$130~Myr,
respectively, for the above \ms\ and \hnb\ setup
(Fig.~\ref{FigDEcc}). While the divergence of trajectories after
$\sim$60~Myr may be attributed to the chaotic 
behavior of the physical system \citep{laskar89,varadi03,
batygin08}, the rapid rise in \DeM\ over only 
$\sim$10~Myr between the 5-Gyr runs with democratic-heliocentric 
(\ms) and Jacobi coordinates (\hnb)
hints to a potential numerical origin.
A rapid \DeM\ rise over $\sim$10~Myr was also observed 
between two \hnb\ runs with identical initial conditions 
and timestep but bodycentric vs.\ Jacobi coordinates.
In addition, orbital solutions obtained with bodycentric 
and Jacobi coordinates yield different Lyapunov times 
for the inner planets (see below).

\ifTWO
\begin{figure}[t]
\def\epsfsize#1#2{0.5#1}
\hspace*{-1.2cm}
\else
\begin{figure}[t]
\def\epsfsize#1#2{0.62#1}
\hspace*{2.5cm}
\fi
\centerline{\vbox{\epsfbox{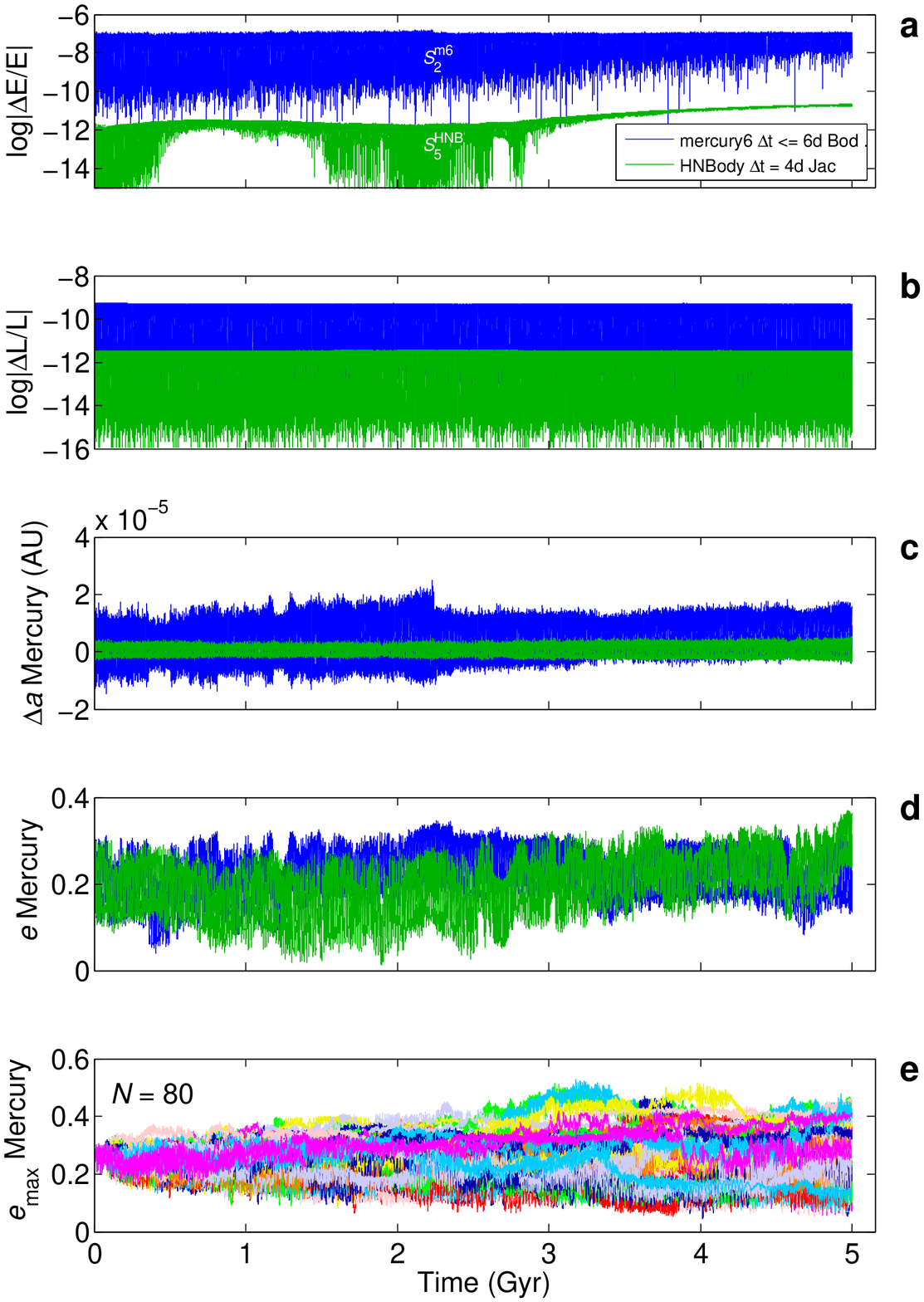}}}
\caption[]{\scs 
Results from 5-Gyr runs with \ms\ and \hnb.
Blue curves (a)$-$(d): \ms, run \#2 (solution $\Sms{2}$), 
$\Dt \leq 6$~days, democratic-heliocentric coordinates.
Green curves (a)$-$(d): \hnb\ run \#5 ($\Shnb{5}$), 
$\Dt = 4$~days, Jacobi coordinates.
(a) Relative energy error, $|\D E/E| = |(E(t)-E_0)/E_0|$. 
(b) Relative angular momentum error, $|\D L/L| = |(L(t)-L_0)/L_0|$. 
(c) Change in Mercury's semi-major axis, $\D a = a(t)-a_0$. 
(d) Mercury's eccentricity (\eM). (e) Mercury's maximum 
eccentricity (per 1~Myr bin)
from all 5-Gyr runs ($N = 80$) with \ms\ and \hnb\ ($N = 40$ 
each). Note time step reduction from~6 to 5~days at 
$\sim$2.24~Gyr in $\Sms{2}$ (see (a) and (c)).\\
}
\label{FigAll5gyr}
\end{figure}

The maximum in the relative energy error in the 5-Gyr 
\ms\ runs was typically $10^4$ times
larger than in \hnb; the corresponding maximum angular 
momentum error about $10^2$ times
larger (Fig.~\ref{FigAll5gyr}).
One might hence be tempted to argue that any differences
between the present 5-Gyr simulations are due to larger 
errors in the \ms\ simulations (less reliable),
rather than heliocentric vs.\ Jacobi coordinates. 
However, this argument is not supported by the results 
of additional ensemble simulations (see below). 
The small offsets in Mercury's initial radial 
distance randomized the initial conditions
and led to complete divergence of eccentricities after
$\sim$100~Myr (Fig.~\ref{FigAll5gyr}). No patterns
in the evolution of Mercury's eccentricity were
observed between adjacent orbits (or sets of orbits);
neither appeared pairs of \ms\ and \hnb\ runs starting
from identical initial conditions lead to correlations 
in the behavior of \eM\ over 5~Gyr. 

However, statistically the 5-Gyr ensemble simulations
with \ms\ (heliocentric coordinates) and \hnb\ (Jacobi 
coordinates) led to significantly different distributions 
for the maximum eccentricities of Mercury (\eMx) achieved
over 5~Gyr ($N=40$ each, Fig.~\ref{FigEmax}). More precisely, 
the null hypothesis that the two \eMx\ samples
from the 5-Gyr \ms\ and \hnb\ ensemble
simulations are random samples from normal distributions 
with equal means can be rejected ($p<0.017$, two-tailed 
Student's $t$-test). Furthermore, the probability for 
\eM\ to grow, say, beyond 0.4 is more than double for the 
5-Gyr \ms\ setup ($11/40=28\%$) than for the \hnb\ setup 
($5/40=13\%$) (Fig.~\ref{FigEmax}).

\ifTWO
\begin{figure}[t]
\def\hsp{-.5cm}
\def\epsfsize#1#2{0.25#1}
\else
\begin{figure}[t]
\def\hsp{2cm}
\def\epsfsize#1#2{0.33#1}
\fi
\centerline{
\hspace*{\hsp}
\epsfbox{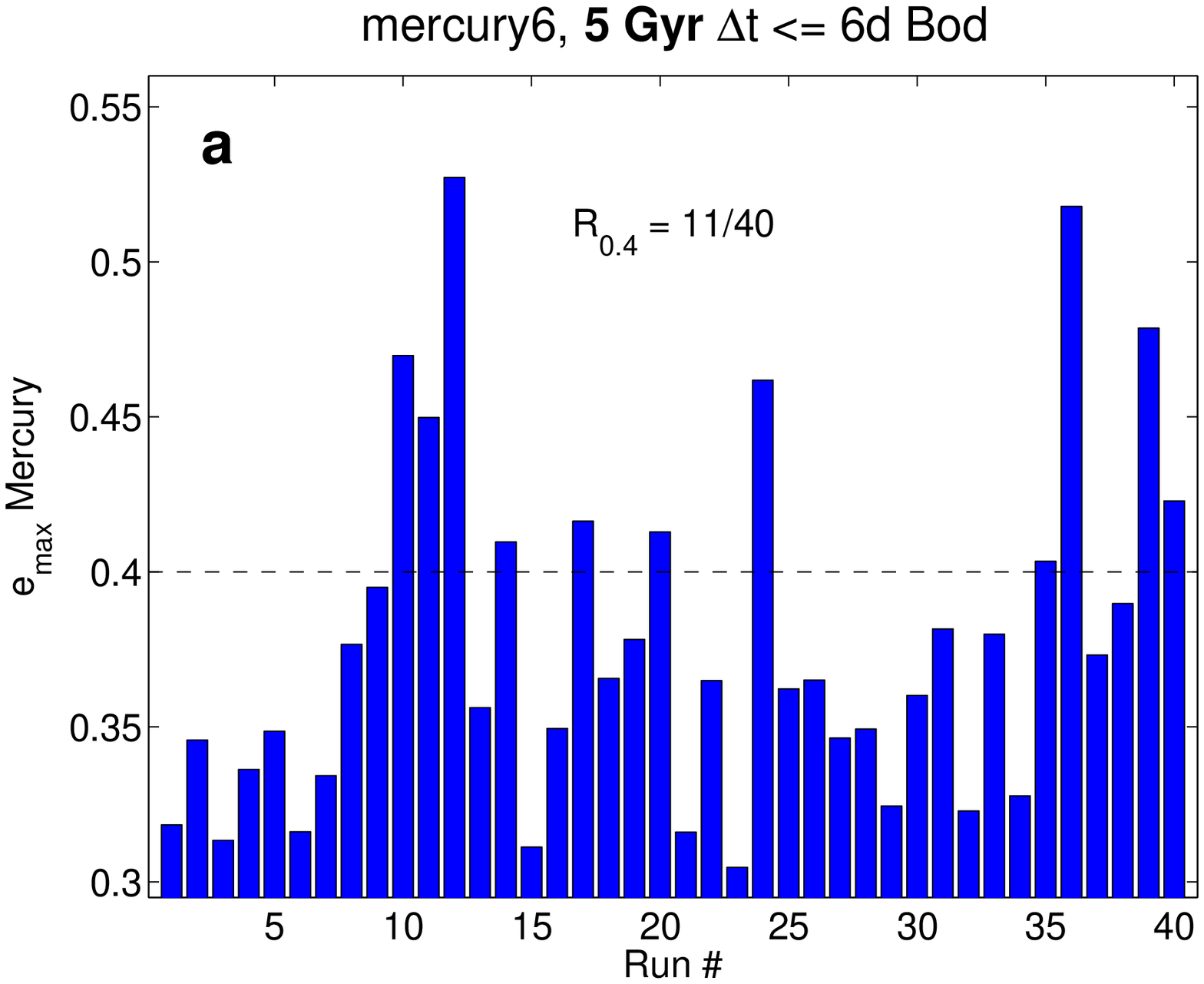}
\epsfbox{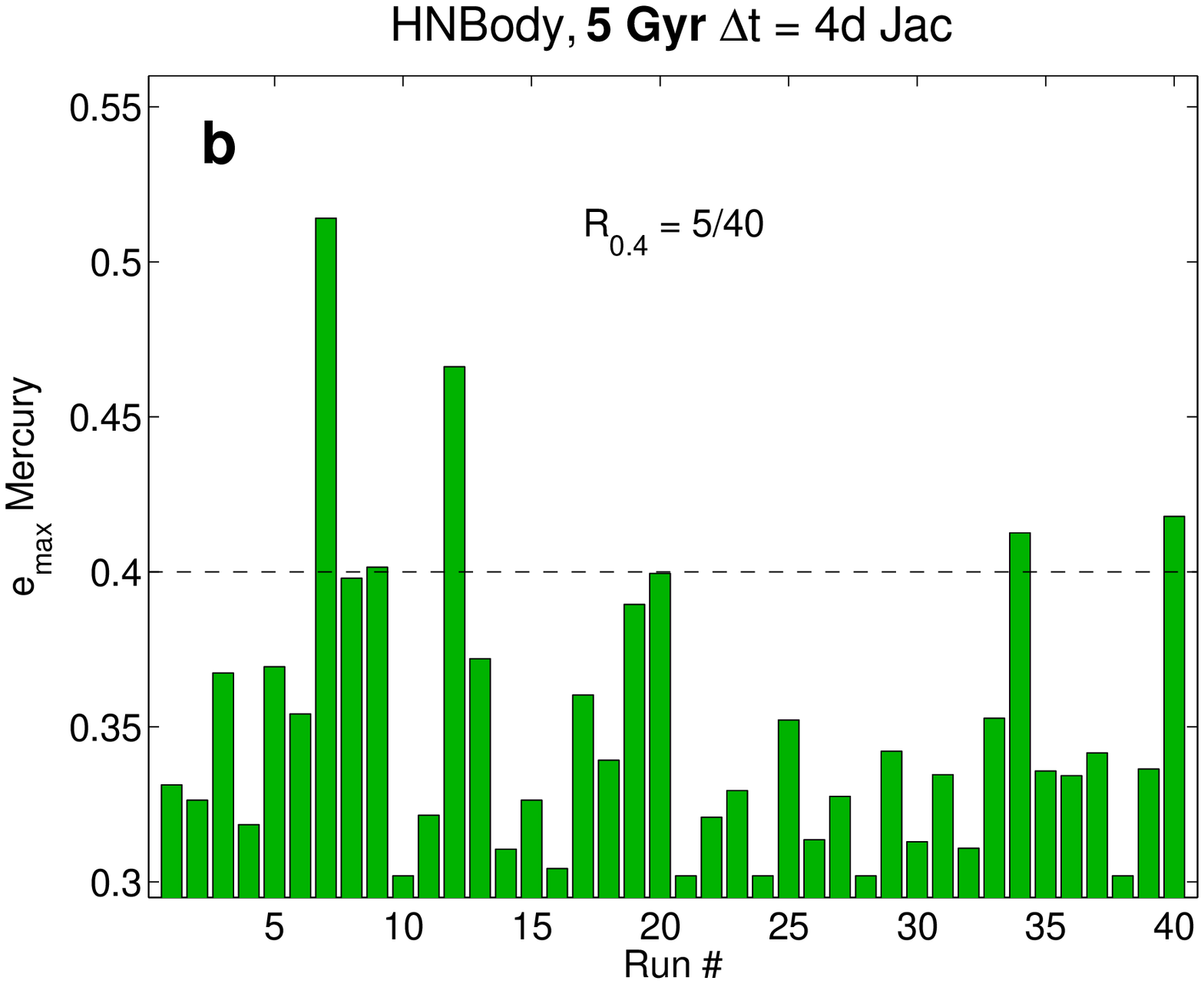}
}
\centerline{
\hspace*{\hsp}
\epsfbox{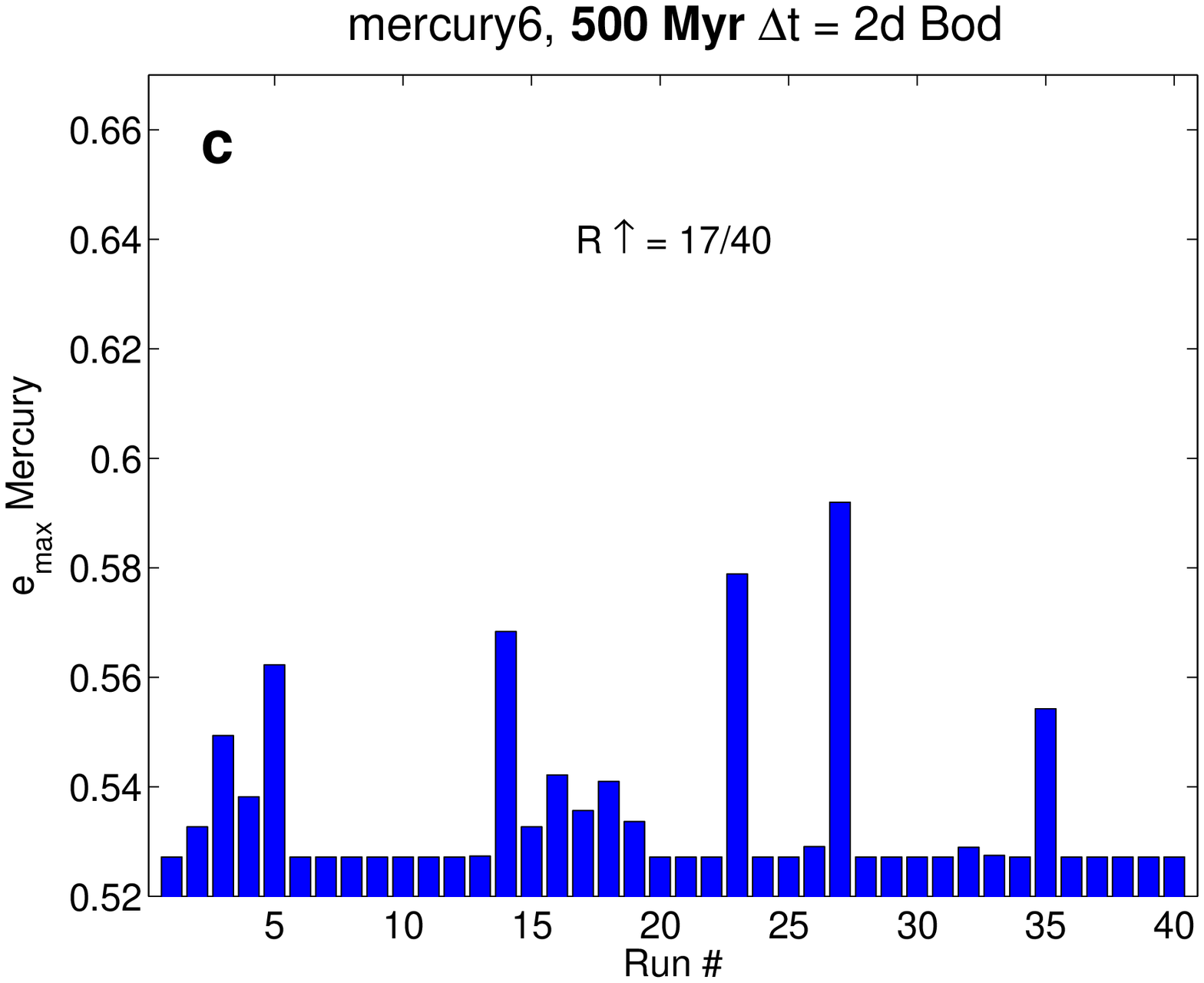}
\epsfbox{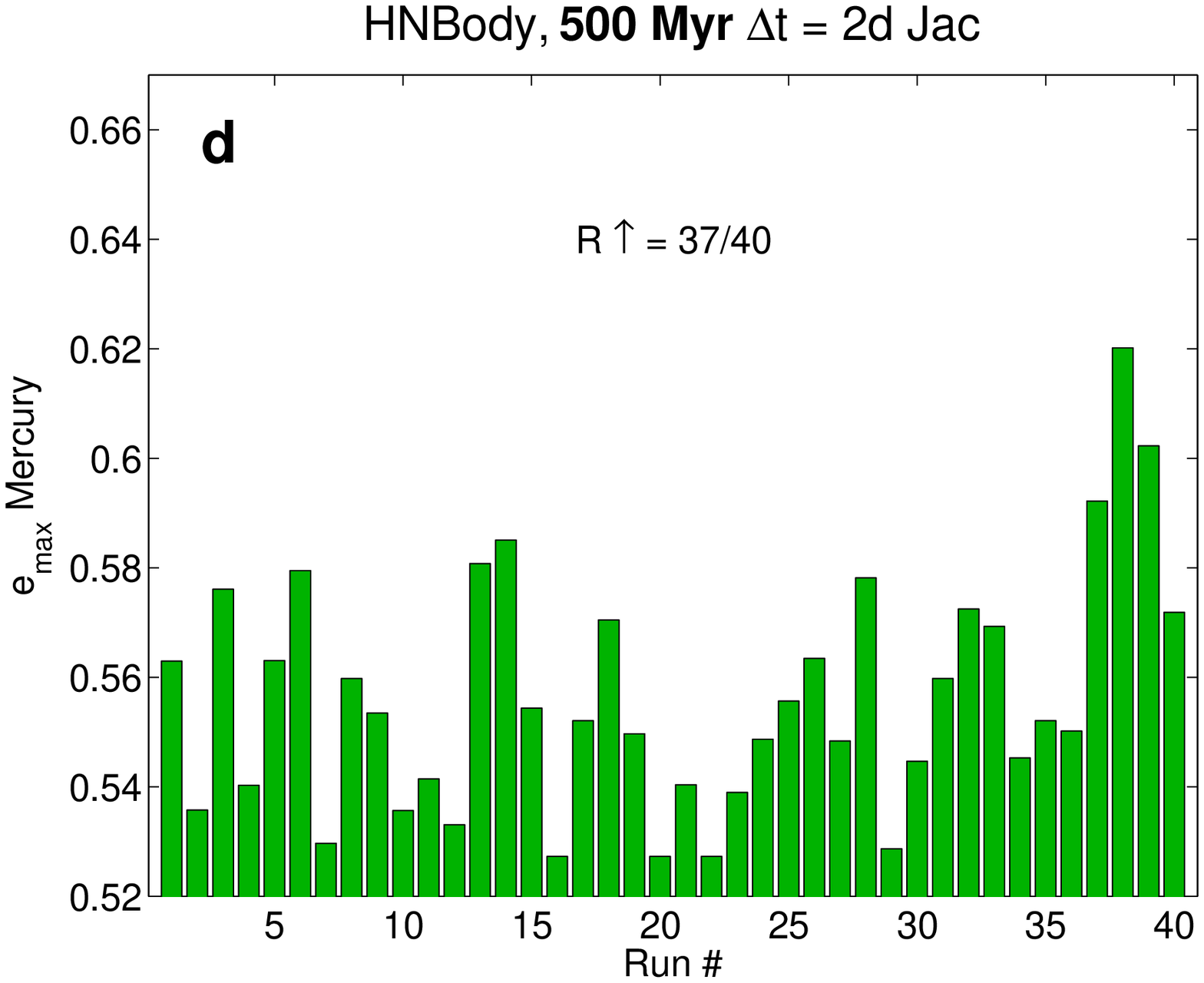}
}
\centerline{
\hspace*{\hsp}
\epsfbox{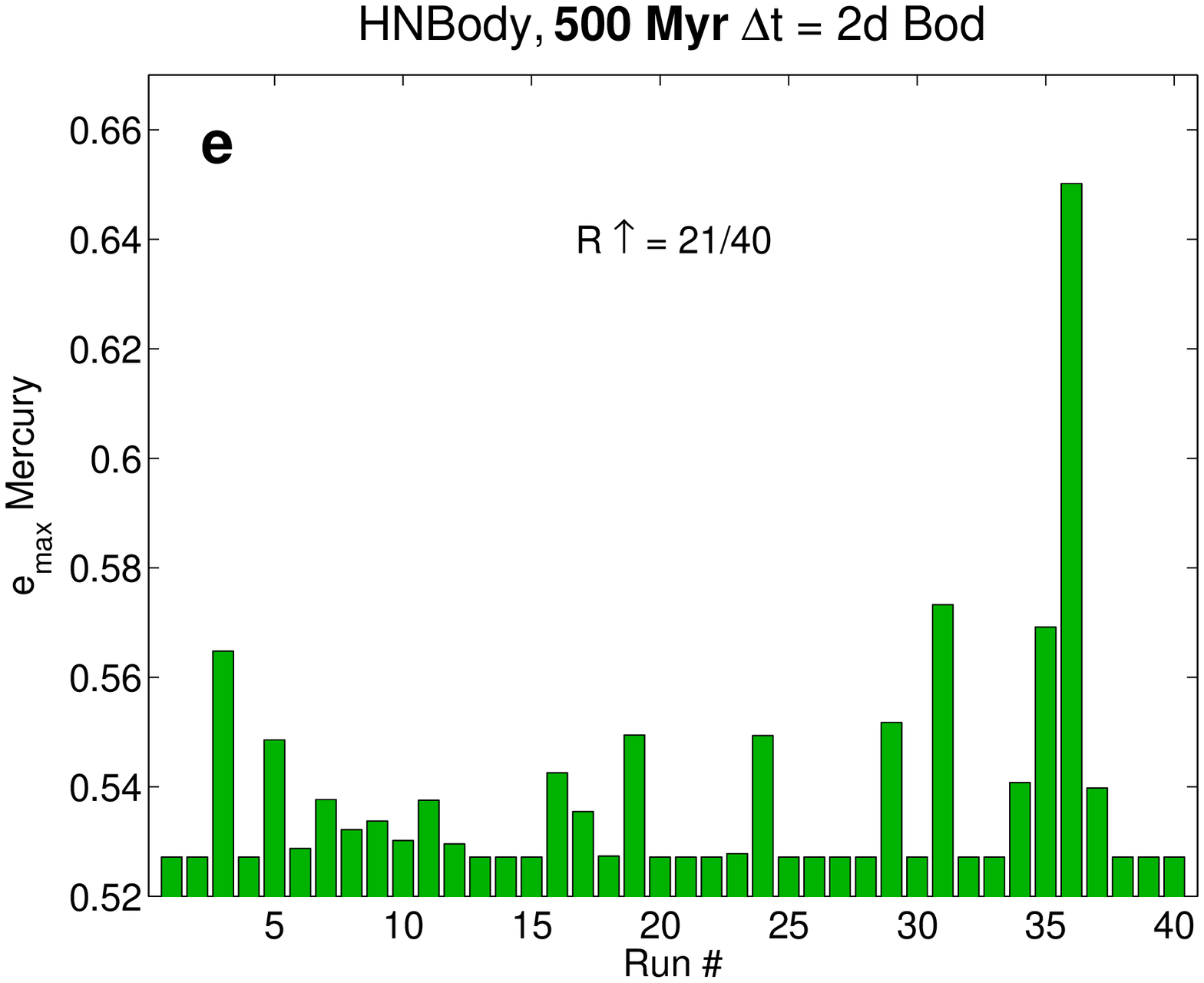}
\epsfbox{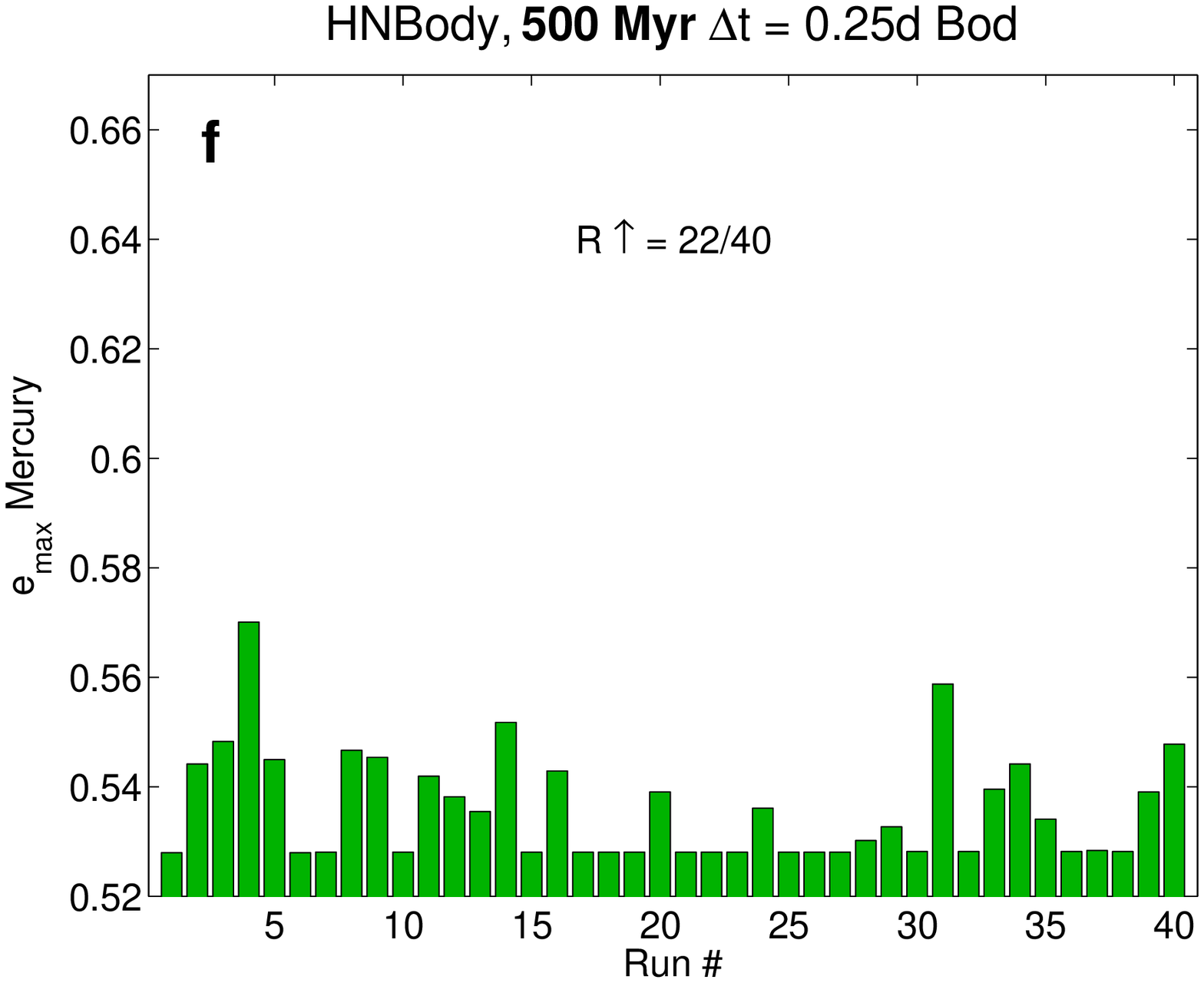}
}
\caption[]{\scs Mercury's maximum eccentricity (\eMx) 
achieved during 5-Gyr and 500-Myr runs with \ms\ (blue bars)
and \hnb\ (green bars). Bod: bodycentric, Jac: Jacobi
coordinates.
(a) Mean $\eMx = 0.377$.
\eM\ grows beyond 0.4 in 11 out of 40 solutions
($R_{0.4}=11/40=28\%$).
(b) Mean $\eMx = 0.349$; $R_{0.4}=5/40=13\%$.
(c) Mean $\eMx = 0.535$; 
\eM\ grows beyond start value of 0.53 in 17 out of 40 solutions
($R_{\up}=17/40=43\%$).
(d) Mean $\eMx = 0.557$; $R_{\up}=37/40=93\%$.
(e) Mean $\eMx = 0.538$; $R_{\up}=21/40=53\%$.
(f) Mean $\eMx = 0.536$; $R_{\up}=22/40=55\%$.
The null hypothesis that two pairwise \eMx\ samples as shown 
in panels $\alpha = a,c,e,f$ and $\beta = b,d$ 
($N = 40$ each) are random 
samples from normal distributions with equal means can be 
rejected at the following significance levels $p_{\alpha\beta}$ 
(two-tailed Student's $t$-test): 
$p_{ab} < 0.017$,
$p_{cd} < 2.2\e{-6}$,
$p_{ed} < 0.0003$,
$p_{fd} < 1.3\e{-6}$.\\
}
\label{FigEmax}
\end{figure}

This statistical discrepancy (obtained at 
relatively low initial $\eM\simeq0.21$) raises 
questions about algorithm performance at high \eM.
For instance, one might expect that \eM\ also grows
larger on average at initially high \eM\ when using 
heliocentric vs.\ Jacobi coordinates. The issue is 
critical for the potential destabilization of the inner 
planets as a result of a large increase in Mercury's
eccentricity. Thus, I conducted additional ensemble 
integrations over 500~Myr starting at the time/conditions 
of the run with the highest $\eM\simeq0.53$ 
(solution \Sms{12}) achieved during the 
5-Gyr simulations (Figs.~\ref{FigEmax} and~\ref{FigResn}). 
The timestep was reduced to 2~days in both the \ms\ 
and \hnb\ setup. Surprisingly, in this case, 
the \ms\ simulations (heliocentric 
coordinates) gave a significantly smaller mean 
\eMx\ value than the \hnb\ simulations (Jacobi coordinates) 
($p<2.2\e{-6}$). Also, the 
probability for \eM\ to increase above the start
value of 0.53 is less than half for the \ms\ 
setup ($17/40=43\%$) than for the \hnb\ setup 
($37/40=93\%$) (Fig.~\ref{FigEmax}). 

\ifTWO
\begin{figure}[t]
\def\epsfsize#1#2{0.5#1} 
\hspace*{0cm}
\else
\begin{figure}[tttttt]
\def\epsfsize#1#2{0.5#1} 
\hspace*{1.5cm}
\fi
\centerline{\vbox{\epsfbox{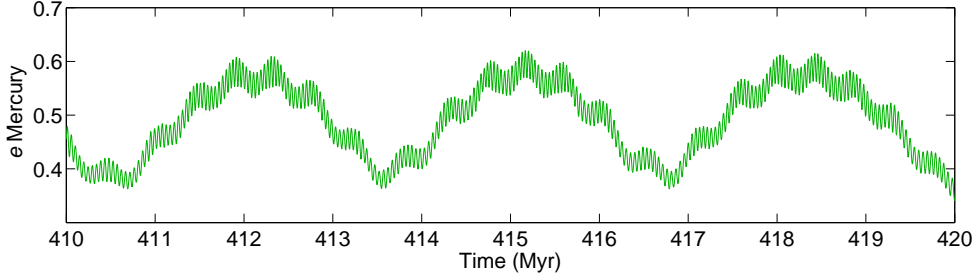}}}
\caption[]{\scs Example of Mercury's eccentricity (\eM) during a
500-Myr run with \hnb\ ($\Dt = 2$~days, Jacobi coordinates).
The pattern is typical during \eM\ increases associated with the 
$(g_1$$-$$g_5)$ resonance (see text).
}
\label{FigResn}
\end{figure}

This result appears unrelated to the integrator package 
but related to the choice of the
integration coordinates. Using \hnb\ with bodycentric 
coordinates for the 500-Myr runs gives a 
mean \eMx\ value similar to the heliocentric \ms\ setup
and significantly lower than the \hnb\ setup with Jacobi
coordinates ($p < 0.0003$, Fig.~\ref{FigEmax}). I also tested 
whether the statistically different \eMx\ at Mercury's
higher initial eccentricity is related to typically larger
errors associated with bodycentric vs.\ Jacobi coordinates
at the same step size. I repeated the 500-Myr runs 
with the bodycentric \hnb\ setup but an eight-fold smaller 
timestep (0.25~d, $\sim$100-fold smaller $|\Delta E/E|$).
However, the basic result remained the same. At high
initial \eM, the setup using bodycentric coordinates leads 
to significantly
smaller mean \eMx\ than the setup using Jacobi
coordinates ($p<1.3\e{-6}$, Fig.~\ref{FigEmax}).
This suggests that the statistical discrepancies are due
to integration coordinates, not errors in energy or
angular momentum (see above).

This poses a quandary because the evolution 
of the physical Solar System, including its probabilistic 
behavior, cannot depend on our choice of coordinate system 
or numerical algorithm. Fundamentally, there is no reason 
to prefer one set of coordinates over the other. 
However, symplectic schemes differ in their implementation
of different integration coordinates (Section~\ref{SecTwoBdy}).
While in symplectic algorithms 
the size of the perturbation term (planet interactions) 
may be numerically larger in heliocentric than in Jacobi 
coordinates, little performance difference has been
reported over 100,000 steps \citep{farres13}. 
Furthermore, heliocentric and Jacobi 
coordinates appear to have opposing effects on \eMx\ at 
low and high initial \eM, pointing to a more complex
origin. 
Importantly, differences per timestep due to integration 
coordinates, which may ultimately cause differences in 
eccentricity, are minuscule.
\ifTWO
\newpage
\vspace*{5cm}
\noindent
\fi
For example, in \hnb\ (\Dt~=~6~days, body vs.\ 
Jacobi) it takes $\sim$200~kyr, or 12~million steps for 
\DeM\ to grow from $10^{-5}$ to $10^{-4}$ ($\sim$1~billion 
steps for Jupiter). Statistical differences between \eMx\ in 
the 500-Myr runs only become significant ($p<0.05$) after 
$\sim$140~Myr, or 25~billion steps (\hnb, \Dt~=~2~days, 
body vs.\ Jacobi).

\section{Fourier analysis: $g_1$$-$$g_5$ Resonance}

Fourier analysis of Mercury's longitude of perihelion
showed that Mercury's eccentricity increase was generally 
associated with a shift in eigenfrequency ($g_1$)
towards Jupiter's forcing frequency ($g_5$). For most 
500-Myr runs, a correlation between \eMx\ and $g_1$ was
observed (Fig.~\ref{FigLPerEcc}).
This is consistent with the pattern of a secular resonance 
involving Mercury and Jupiter (plus Venus' participation)
\citep{batygin08,laskar08,lithwick11,boue12}
and appears to be the common cause for \eM\ increases
in nearly all 500-Myr runs, regardless of integrator 
package or coordinates. Note that contributions from 
general relativity \citep{einstein16} as included here 
\citep{mikkola98,soffel89} are important as they 
universally move $g_1$ up (by $\sim$0.43''~y\pmo\ 
at present)
and away from the $(g_1$$-$$g_5)$ resonance, 
substantially reducing the probability for large \eM\ 
increases across all simulations \citep{laskar08,laskar09}
(Fig.~\ref{FigLPerEcc}).

\ifTWO
\begin{figure}[t]
\def\epsfsize#1#2{0.52#1}
\hspace*{-1cm}
\else
\begin{figure}[t]
\def\epsfsize#1#2{0.7#1}
\hspace*{2cm}
\fi
\centerline{\vbox{\epsfbox{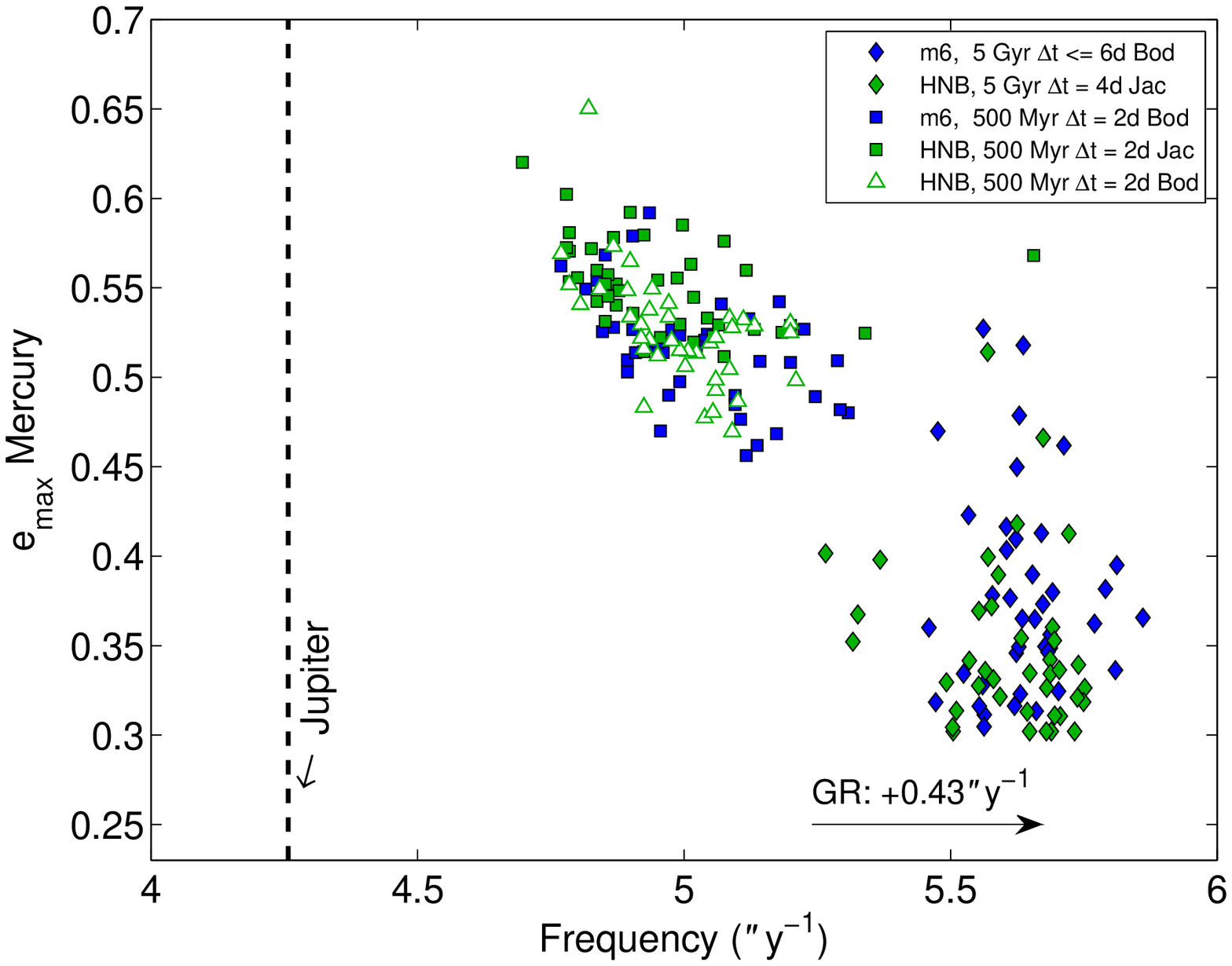}}}
\caption[]{\scs Mercury's maximum eccentricity (\eMx) vs.\ 
eigenfrequency $g_1$ in arcsec y\pmo\ (symbols; $g_1$ 
from Fourier analysis of Mercury's longitude of perihelion,
$\varpi_1$). m6: \ms, HNB: \hnb, Bod: bodycentric, Jac: Jacobi
coordinates.
Dashed line: Jupiter's forcing frequency. For the 5-Gyr
runs, \eMx\ and $g_1$ values shown were determined using 
output for \eM\ and $\varpi_1$ over the full 5-Gyr time 
interval. For the 500-Myr runs, \eMx\ and $g_1$
values shown
were determined only over the final 250-Myr interval 
of the runs. This improves the representation of solutions
with long-term decline in \eM\ below the 0.53-start value. 
Otherwise these solutions would all plot at constant 
$\eMx = 0.53$ despite large variations in $g_1$.
Contributions from general relativity (included in 
all simulations) universally move $g_1$ up (by 
$\sim$0.43''~y\pmo\ at present, arrow) and away from the 
$(g_1$$-$$g_5)$ resonance.
}
\label{FigLPerEcc}
\end{figure}

More importantly, in the 500-Myr runs bodycentric 
coordinates showed on average a higher tendency towards
resonance damping and hence larger $g_1$
and smaller \eMx, compared to Jacobi coordinates.
This tendency is significant
and roughly halves the probability for \eM\ to increase 
beyond 0.53 in the 500-Myr runs (Fig.~\ref{FigEmax}).
The cause for the statistical differences originating 
from different integration coordinates as found here is 
not obvious, neither is clear which (if any) 
of the methods provides accurate probability predictions
over $10^8$--$10^9$-year timescale (see below). 

\section{Mercury's eccentricity and Lyapunov times 
from different simulations}

Starting from slightly different initial conditions 
(2.73~m offset in Mercury's initial radial distance), 
the computed difference in Mercury's eccentricity (\DeM)
between two solutions grows slowly over the first
50~Myr or more when using the same program and the
same integration coordinates
(only bodycentric or only Jacobi coordinates, 
Fig.~\ref{FigDEcc}a--e). This is also true for
the same initial conditions, different programs (\ms\ vs.\ 
\hnb) and same integration coordinates (both
bodycentric, Fig.~\ref{FigDEcc}f). 
However, using the same initial conditions but
different integration coordinates (bodycentric vs.\
Jacobi coordinates), \DeM\ between two solutions grows 
quickly over the first 10~Myr (Fig.~\ref{FigDEcc}g--j).
This feature was observed irrespective of whether the
same or different programs were used, whether general
relativity contributions were included or not, and whether
the initial eccentricity was low or high (initial 
$\eM = 0.21/0.53$, see above).

During the slow initial rise (Fig.~\ref{FigDEcc}a--f),
\DeM\ is usually dominated by polynomial growth and may
increase linearly with time in a log-log 
plot \citep{laskar89,varadi03}.
The subsequent rapid rise after $\gtrsim 50$~Myr until 
$\DeM \simeq \eM$ has been attributed to the 
chaotic nature of the physical system 
\citep{laskar89,varadi03,batygin08}. However, it is not
clear why the system's chaotic behavior starts dominating 
\eM's evolution after e.g.\ $\sim$60~Myr in \ms\ with
bodycentric coordinates (Fig.~\ref{FigDEcc}a) 
but only after $\sim$120~Myr in \hnb\ with Jacobi 
coordinates (Fig.~\ref{FigDEcc}c). The initial $\DeM(t=0)$ 
in both the \ms\ and \hnb\ simulations was 
$\sim$$3\e{-11}$. The time evolution of the absolute 
\DeM\ maxima of the two curves (Fig.~\ref{FigDEcc}a,c) may be
fit to simple functions assuming linear and exponential 
growth of the initial \DeM\ (Fig.~\ref{FigDeMfit}). The 
exponential function fits the rapid growth phase well 
in the \ms\ simulations (Body-Body) with an estimated 
Lyapunov time $\tau \simeq 3.2$~Myr. However, the 
exponential function is a poor fit to the \hnb\ 
simulations (Jacobi-Jacobi) with an estimated Lyapunov 
time $\tau \simeq 6$~Myr (Fig.~\ref{FigDeMfit}). 

Nevertheless,
these estimated Lyapunov times are in good agreement with 
estimates of Lyapunov exponents from phase space separation 
of two nearby orbits over time (Fig.~\ref{FigLyap}).
\tcx{Note that strictly, Lyapunov exponents are derived
from the solution of the variational equations,
rather than from the evolution of two system trajectories
\citep[e.g.][]{holman96,tancredi01,morbidelli02}.
However, the emphasis here is on the {\it relative} difference
between simulations with, say, different integration 
coordinates, while applying the same method for estimating
Lyapunov exponents. Importantly, {\it absolute} estimates obtained
here are consistent with results from previous 
studies \citep{laskar89,varadi03,batygin08}.}

\ifTWO
\onecolumn
\begin{figure}[p]
\def\epsfsize#1#2{0.44#1}
\else
\begin{figure}[t]
\def\epsfsize#1#2{0.44#1}
\fi
\centerline{\vbox{
\epsfbox{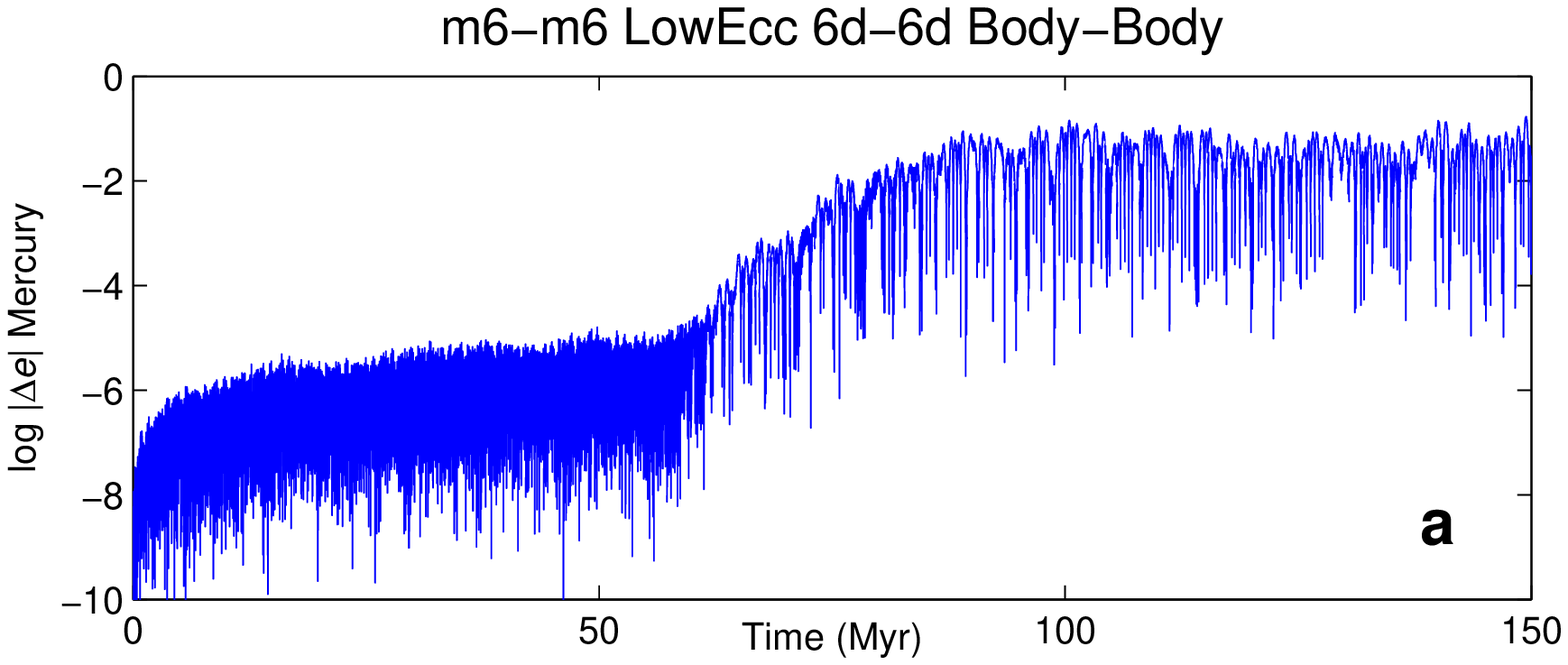}      
\epsfbox{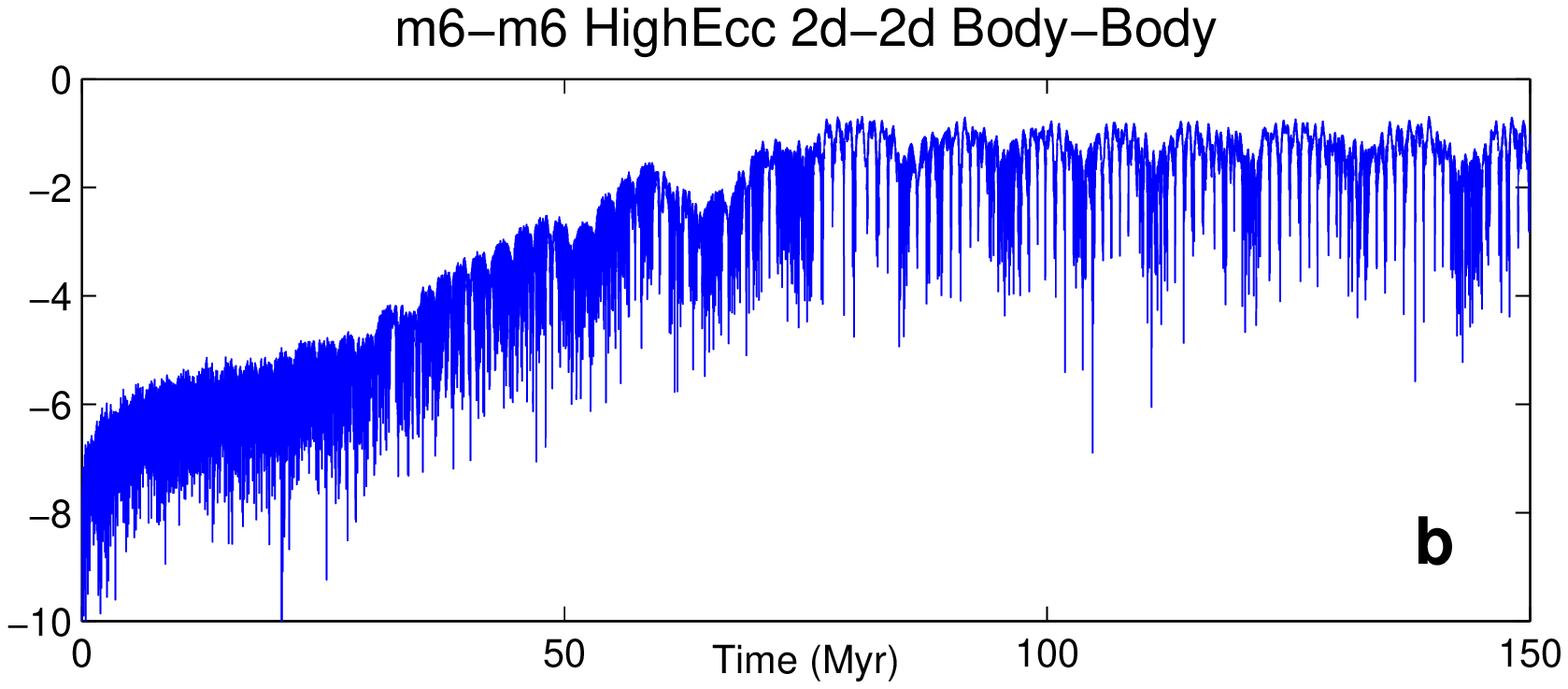}     
}}
\centerline{\vbox{
\epsfbox{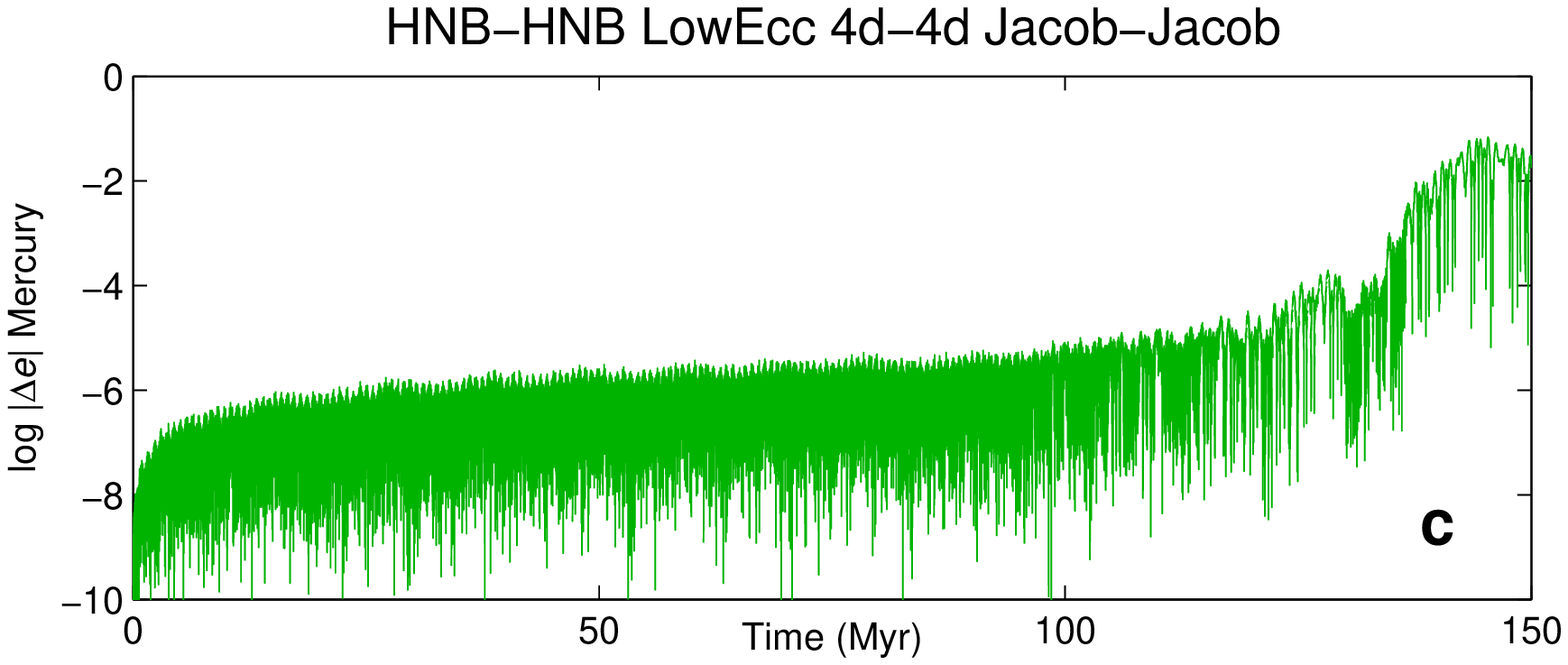}    
\epsfbox{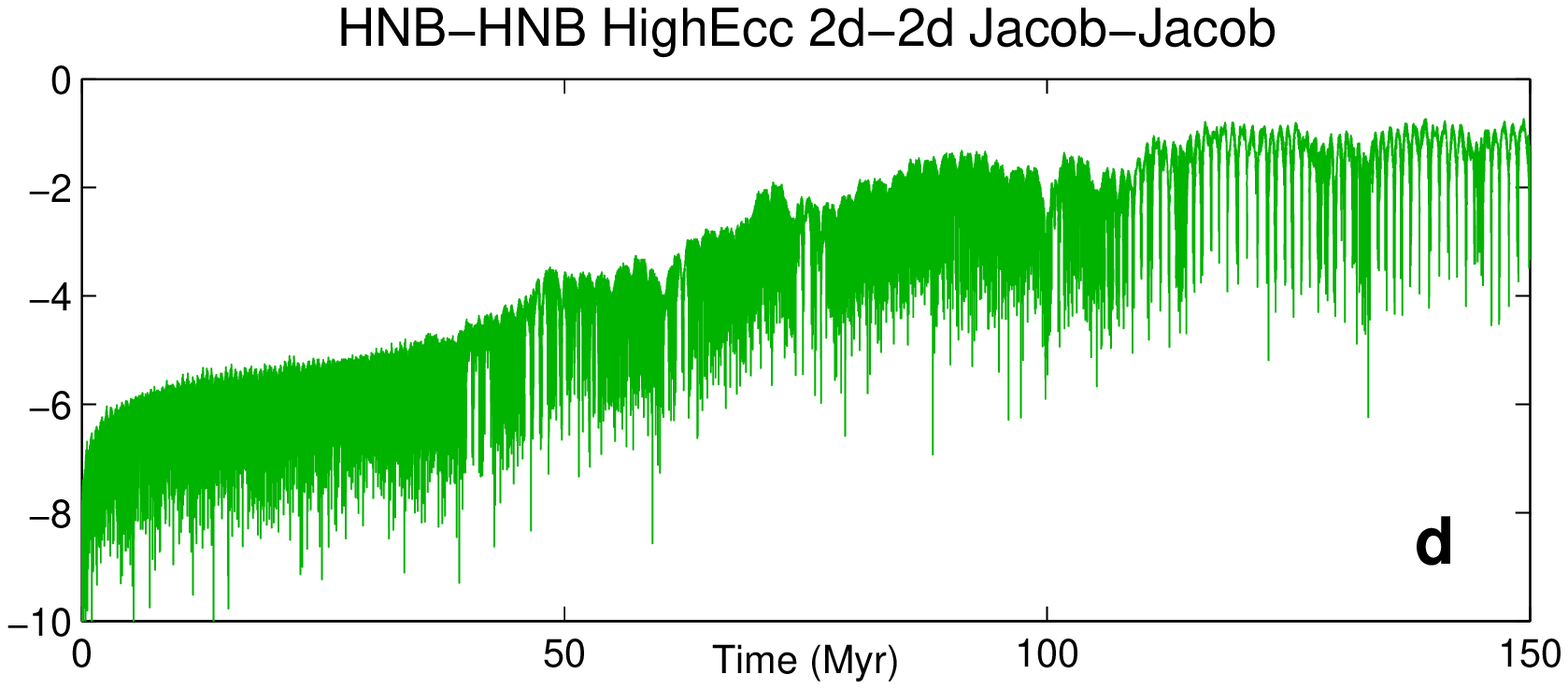}   
}}
\centerline{\vbox{
\epsfbox{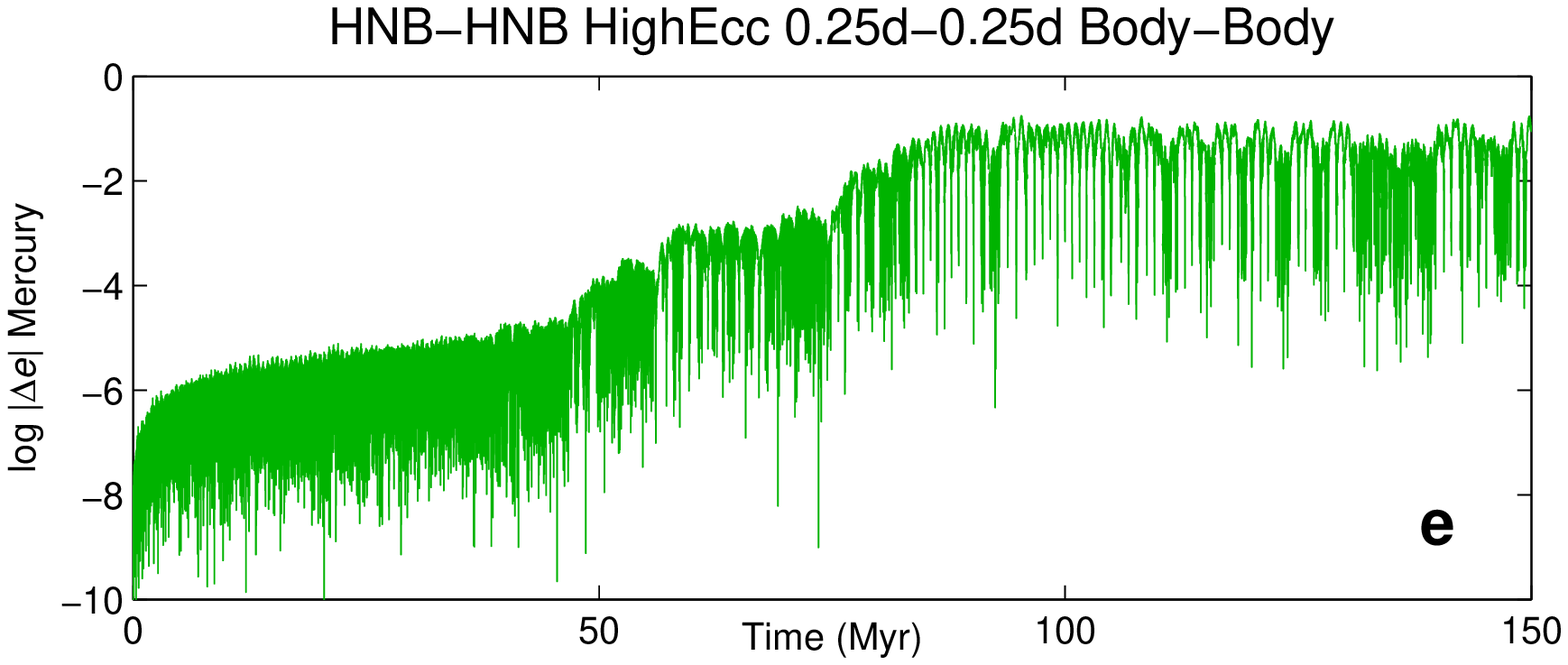} 
\epsfbox{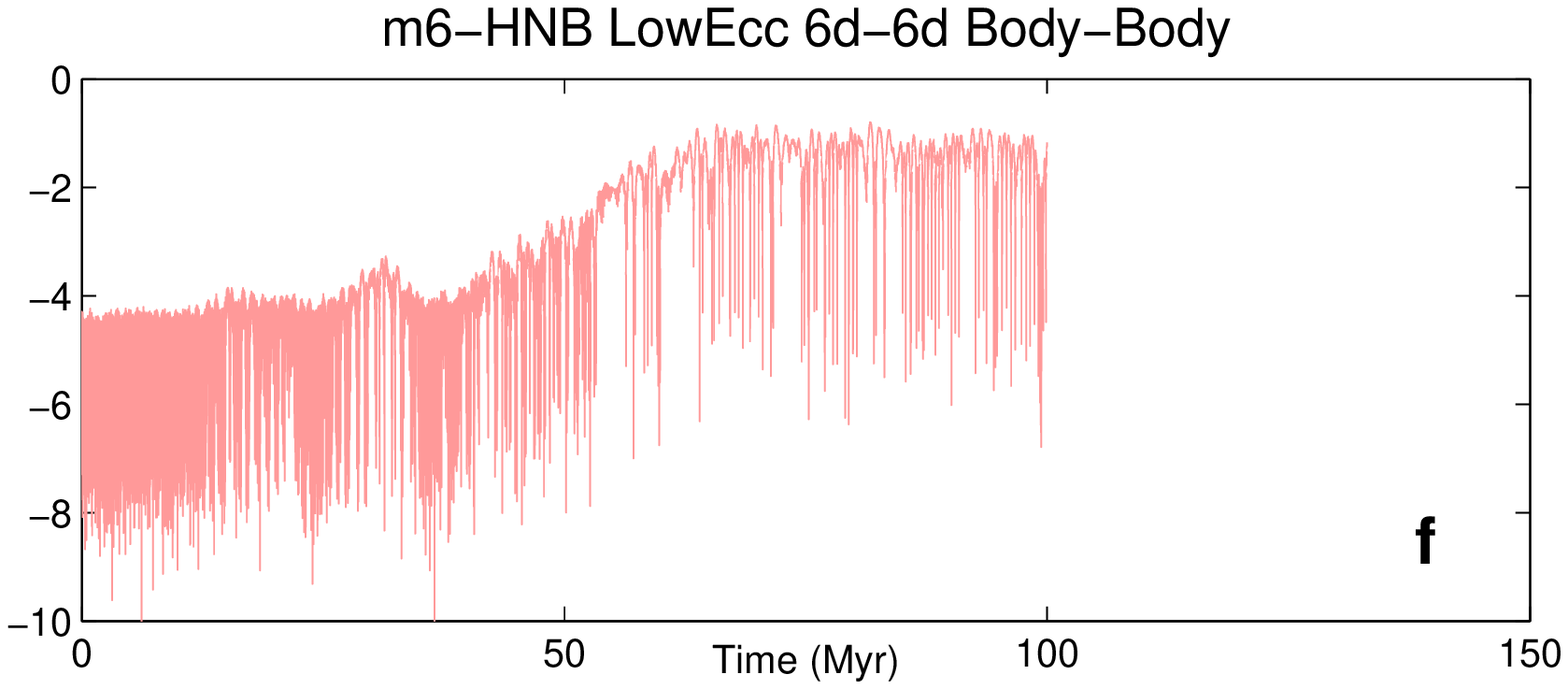}     
}}
\centerline{\vbox{
\epsfbox{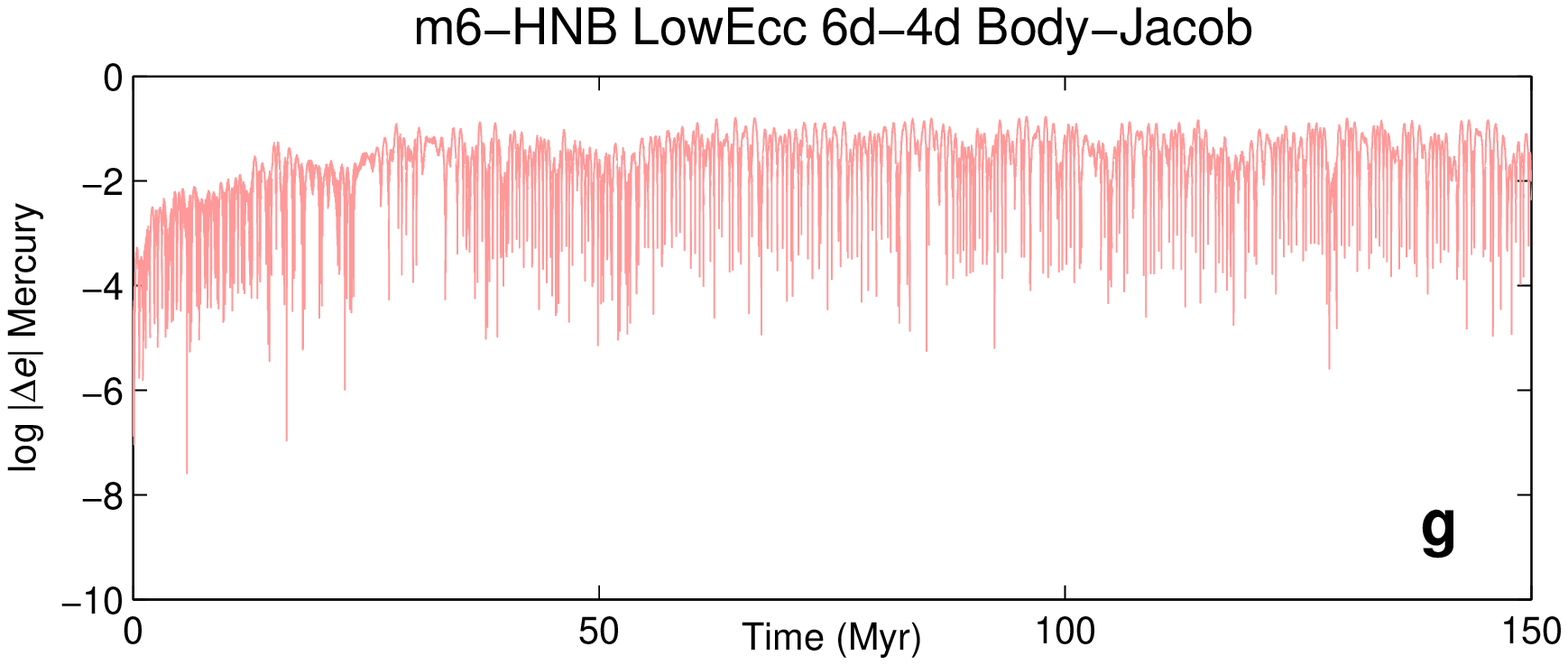}     
\epsfbox{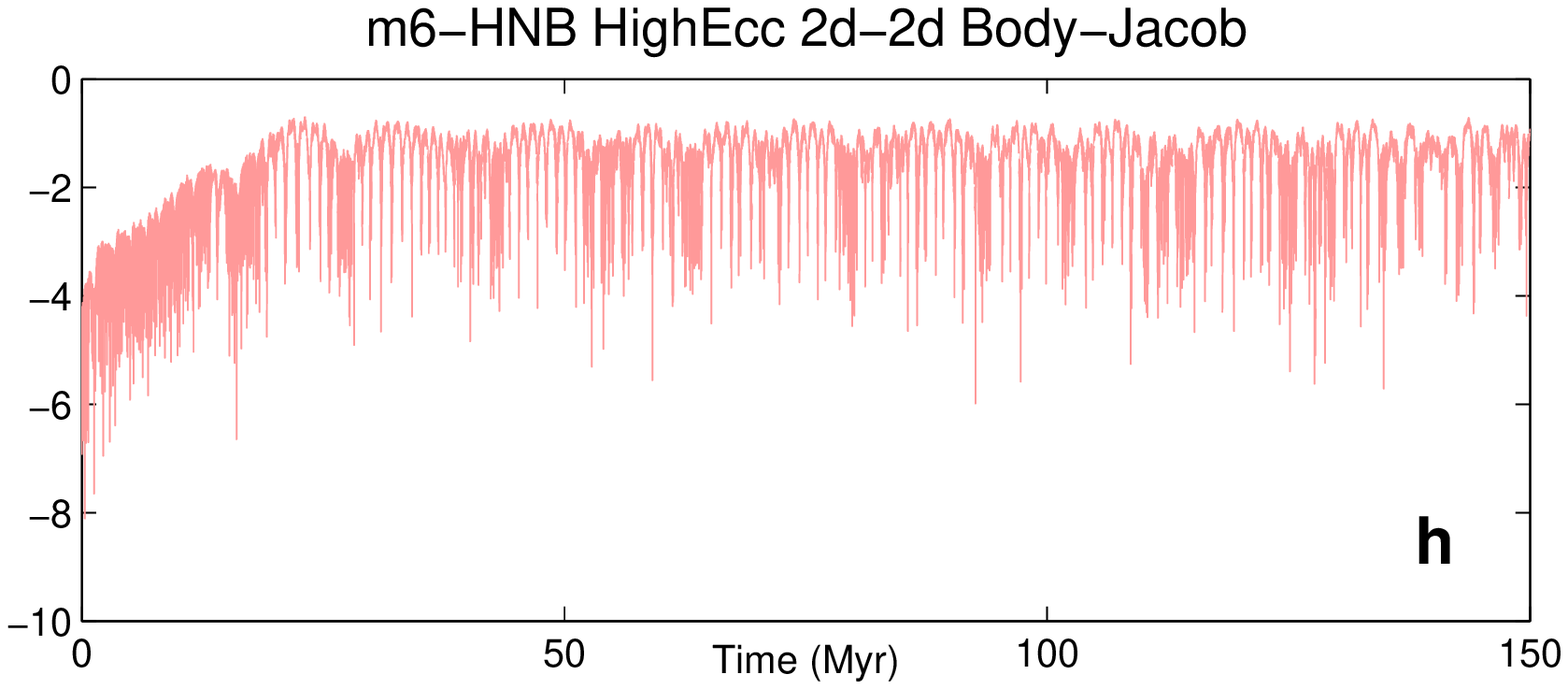}    
}}
\centerline{\vbox{
\epsfbox{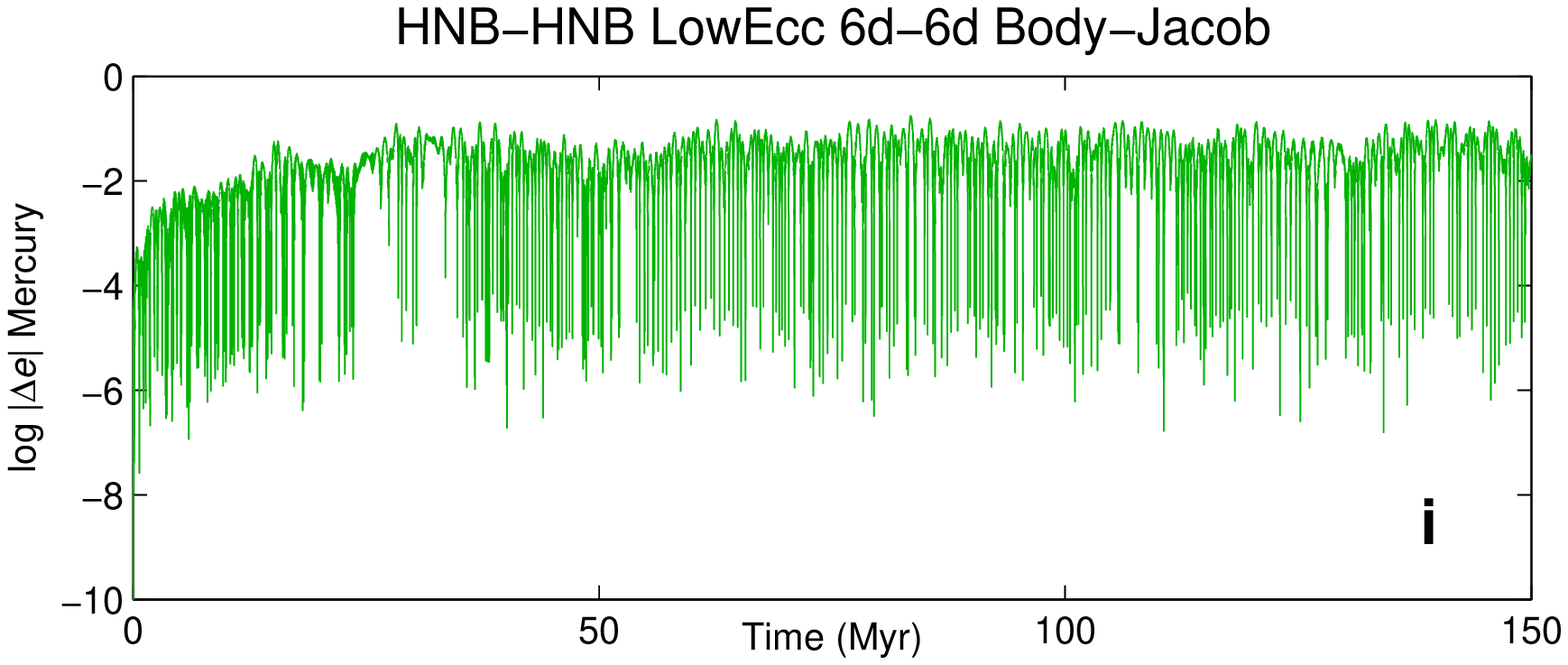}      
\epsfbox{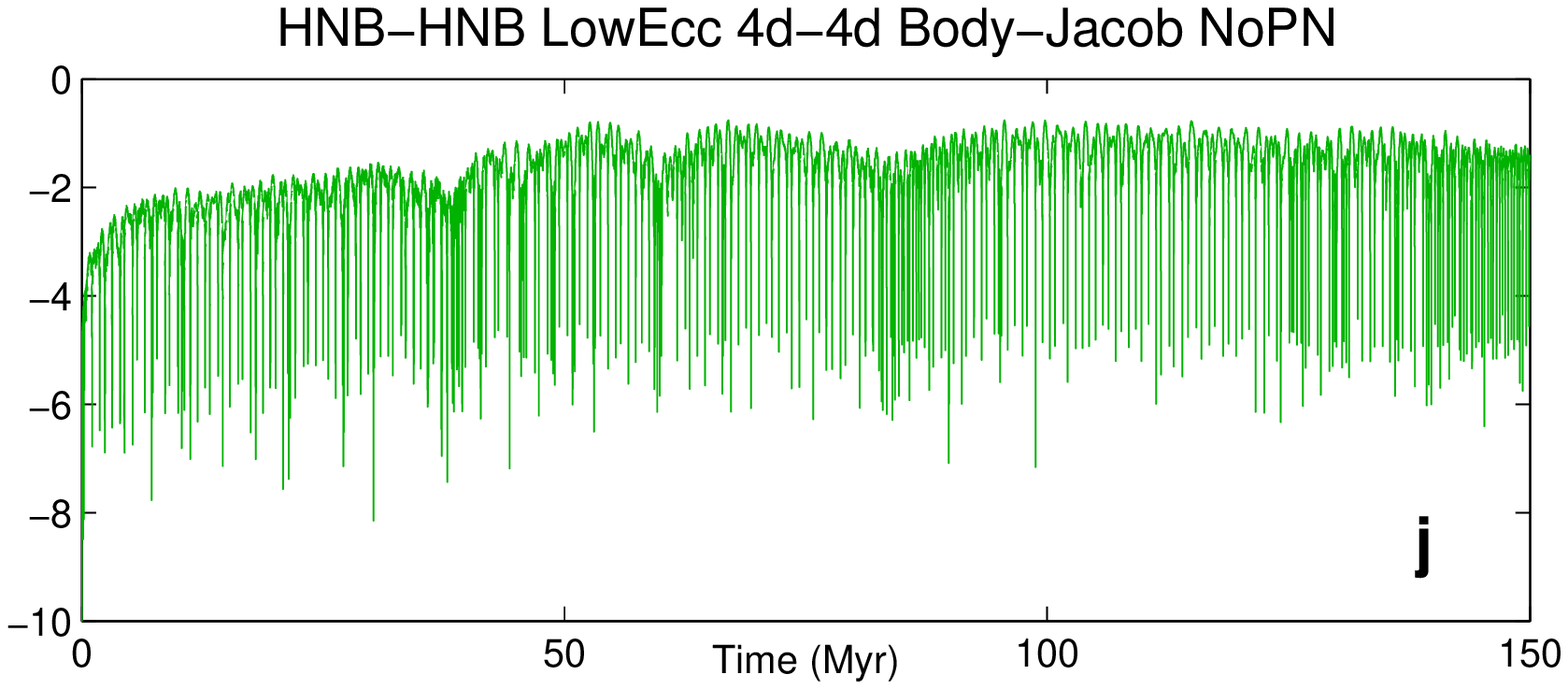} 
}}
\caption[]{\scs 
Computed difference in Mercury's eccentricity (\DeM) between
two runs each per panel for different integrations and options. 
Blue curves: both runs with \ms\ (m6), 
green curves: both runs with \hnb\ (HNB),
light red curves: one run with \ms, the other with \hnb.
Body: bodycentric, Jacob: Jacobi coordinates. $m$d--$n$d 
indicates the timestep in both runs (in days).
LowEcc/HighEcc: initial $\eM = 0.21/0.53$ (see text).
NoPN: No contributions from general relativity. Panels
a--e show differences in \eM\ between solutions with slightly
different initial conditions (2.73~m offset in Mercury's initial
radial distance); panels f--j: identical initial 
conditions. Note rapid \DeM\ rise in g--j 
(all Body-Jacobi).
}
\label{FigDEcc}
\end{figure}
\ifTWO\twocolumn\fi

\ifTWO
\begin{figure}[tttttt]
\def\epsfsize#1#2{0.5#1}
\hspace*{-.5cm}
\else
\begin{figure}[tttttt]
\def\epsfsize#1#2{0.45#1}
\hspace*{3cm}
\fi
\centerline{\vbox{
\epsfbox{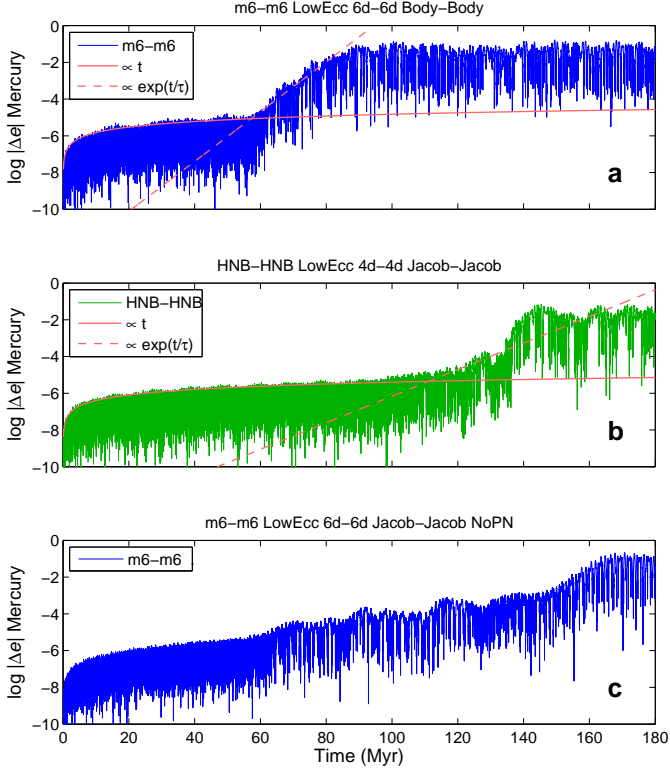}
}}
\caption[]{\scs
Computed difference in Mercury's eccentricity (\DeM) between
two runs each per panel with slightly different initial 
conditions (2.73~m offset in Mercury's initial radial distance)
for different integrations and options. 
Blue curves: both runs with \ms\ (m6), 
green curve: both runs with \hnb\ (HNB),
red curves: simple fit functions assuming linear and exponential 
growth of initial \DeM.
Body: bodycentric, Jacob: Jacobi coordinates. $m$d--$n$d 
indicates the timestep in both runs (in days).
LowEcc: initial $\eM = 0.21$ (see text).
NoPN: No contributions from general relativity. 
Note different slopes (estimated Lyapunov 
time $\tau$) of exponential fits in (a) and (b)
(red dashed lines; linear on logarithmic y-scale).
}
\label{FigDeMfit}
\end{figure}

The Lyapunov exponents were determined from two runs each
with \ms\ and \hnb\ (one fiducial, one shadow orbit each) 
without renormalization
and an initial separation of $5\e{-14}$~AU in Mercury's 
$x$-coordinate. Renormalization was tested but yielded spurious 
results as reported before \citep{tancredi01}. If during
each time interval $\dt$, the distance in phase space
grows exponentially, then $d_j = d^0_j \exp(\gamma \cdot \dt)$,
where $d^0_j$ and $d_j$ is the initial and final distance.
An average $\gamma_k$ after $t = k \cdot \dt$ may then 
be computed as:
\beqn
\gamma_k = \q{1}{k \cdot \dt} \sum_{j=1}^k \ \ln (d_j/d^0_j) \ ,
\eeqn
where $\dt$ was set to 20~kyr. A log-log plot of $\gamma_k$ vs.\
time yields a straight line with negative constant slope 
until an inflection point is reached after which $\gamma_k$ 
approaches a non-zero value (Fig.~\ref{FigLyap}).
The inflection point corresponds to the time in the integration 
where exponential growth starts dominating the evolution
of the phase space distance between the two orbits.
Note that the graphs in Fig.~\ref{FigLyap} have been 
truncated shortly after the inflection point to emphasize
the plateau in $\gamma_k$ ($\gamma_k$ keeps decreasing 
subsequently as $d$ is bounded without renormalization,
not shown).
The corresponding Lyapunov exponents for the \ms\ and
\hnb\ standard runs are $10^{-6.44}$ and $10^{-6.74}$ 
(i.e.\ Lyapunov times of 2.8~Myr and 5.5~Myr), respectively,
in good agreement with the Lyapunov times
estimated from differences in Mercury's eccentricity 
evolution (Fig.~\ref{FigDeMfit}). Note that based on 
different approaches it may be difficult to measure Lyapunov 
times with an {\it absolute} accuracy much better than a 
factor of two \citep{murray99}. However, based on
the current approach (simulations with identical initial 
conditions, etc.), it is not difficult to measure Lyapunov 
times with a {\it relative} accuracy better than a 
factor of two (Figs.~\ref{FigDeMfit} and~\ref{FigLyap}).

\ifTWO
\begin{figure}[tttttt]
\def\epsfsize#1#2{0.42#1}
\hspace*{-.3cm}
\else
\begin{figure}[tttttt]
\def\epsfsize#1#2{0.55#1}
\hspace*{3cm}
\fi
\centerline{\vbox{
\epsfbox{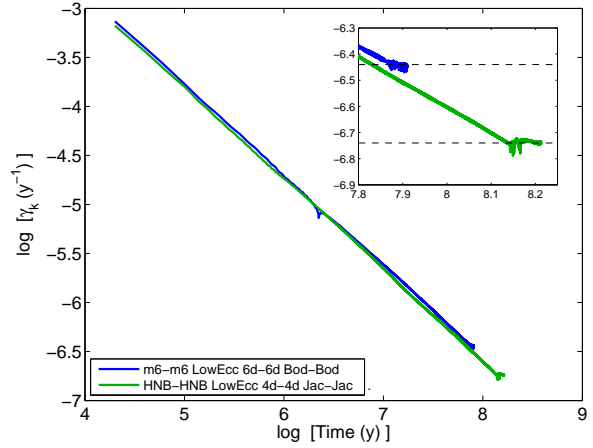}
}}
\caption[]{\scs
Lyapunov exponents estimated from phase space separation 
of two nearby orbits over time. Blue curve: both runs with 
\ms\ (m6), green curve: both runs with \hnb\ (HNB).
Graphs have been truncated shortly after the inflection 
point (see text). Bod: bodycentric, Jac: Jacobi coordinates. 
$m$d--$n$d  indicates timestep in both runs (in days).
LowEcc: initial $\eM = 0.21$.
}
\label{FigLyap}
\end{figure}

I also
computed \DeM\ over time for two simulations with
slightly different initial conditions using
Jacobi coordinates in \ms\ (mixed-variable symplectic
algorithm) without contributions
from general relativity (Fig.~\ref{FigDeMfit}). In
this case, selecting parameters for an exponential fit 
is not obvious  but it is clear that it takes more than
160~Myr for \DeM\ to approach the magnitude of \eM.
The bottom line is that the estimated Lyapunov 
times for Mercury's orbit from the standard 
\ms\ and \hnb\ integrations (Figs.~\ref{FigDeMfit} 
and~\ref{FigLyap}) differ at least by a factor of 
two, which is easily measurable given the relative 
accuracy of the current approach. However, the 
chaotic behavior of the Solar System is a physical 
property that cannot depend on the 
numerical algorithm chosen to describe the system. 
Yet the results presented here suggest that an exponential
divergence of trajectories starts dominating 
the numerical solutions after significantly different time 
intervals, depending on integration coordinates
(Figs.~\ref{FigDEcc}, \ref{FigDeMfit}, \ref{FigLyap}). 
In addition, the rapid \DeM\ growth over the first 
10~Myr between two solutions using the same initial 
conditions but different integration coordinates 
(Fig.~\ref{FigDEcc}g--j) suggests a fundamental 
difference between algorithms using bodycentric and 
Jacobi coordinates.

\section{Algorithm tests}

The results presented above raise questions about 
numerical algorithm performance and the use of integration 
coordinates. Why do heliocentric and Jacobi coordinates lead 
to statistically different results? Which algorithm/integration 
coordinates provide more accurate results? This
section describes several tests aiming at resolving the 
dilemma of the dependency of statistical results on integration 
coordinates.

\subsection{2-Body problem: Position errors \label{SecTwoBdy}}

One approach for testing the accuracy of a numerical 
algorithm is to study the 2-body problem, for which an 
analytical solution exist (strictly, Kepler's equation 
is solved numerically though). While substantially less 
complex than the general $n$-body problem, in the
following the 2-body 
problem will illustrate one important difference between 
symplectic algorithms with heliocentric and Jacobi 
coordinates.

The Hamiltonian for the 2-body problem may be written as:
\beqn
H = \q{|\v{p_0}|^2}{2 m_0} + \q{|\v{p_1}|^2}{2 m_1} 
    - \q{G m_0 m_1}{|\v{x}_0 - \v{x}_1|} \ ,
\eeqn
where $G$ is the gravitational constant and
$\v{p}_i$, $m_i$, and $\v{x}_i$ are the momenta,
masses, and positions of the two bodies. The problem
can be simplified via canonical transformations, using 
e.g.\ democratic-heliocentric 
(DH) and Jacobi coordinates, denoted here as:
\beq
(\v{x}  \ , \ \v{p} ) & \lrarr 
(\v{\Q} \ , \ \v{\P}):   \  & \mbox{Democratic-Heliocentric} \\
(\v{x}  \ , \ \v{p} ) & \lrarr 
(\v{Q}  \ , \ \v{P} ):   \  & \mbox{Jacobi} \ .
\eeq

In DH coordinates, $\v{\Q}_0$ is the position of the center 
of mass and $\v{\Q}_1$ is the heliocentric position
of $m_1$. One may use a generating function 
$F_3(\v{\Q},\v{p})$ of the new positions and the old 
momenta \citep{duncan98}:
\beqn
F_3 = & & - \v{p}_0 (\v{\Q}_0 - \q{m_1}{M} \v{\Q}_1) \nn \\
      & & - \v{p}_1 (\v{\Q}_0 - \q{m_1}{M} \v{\Q}_1 + \v{\Q}_1) \ ,
\eeqn
where $M = m_0 + m_1$.
The relationship between old and new variables then is:
\beq
-\q{\prt F_3}{\prt \v{p}_0}  & = & \v{x}_0 
    = (\v{\Q}_0 - \q{m_1}{M} \v{\Q}_1)           \\
-\q{\prt F_3}{\prt \v{p}_1}  & = & \v{x}_1 
   = (\v{\Q}_0 - \q{m_1}{M} \v{\Q}_1 + \v{\Q}_1) \\
-\q{\prt F_3}{\prt \v{\Q}_0} & = & \v{\P}_0 
   =  \v{p}_0 + \v{p}_1                          \\
-\q{\prt F_3}{\prt \v{\Q}_1} & = & \v{\P}_1
   =  \v{p}_1 - \q{m_1}{M} ( \v{p}_0 + \v{p}_1) \ .
\eeq
Note that
$\v{\P}_0$ is the total momentum and $\v{\P}_1$ is the 
barycentric momentum of $m_1$. After some algebra,
the Hamiltonian becomes:
\beqn
H =   \left(
      \q{|\v{\P}_1|^2}{2 m_1} 
    - \q{G m_0 m_1}{|\v{\Q}_1|} \right)
    + \q{|\v{\P}_0|^2}{2 M}
    + \q{|\v{\P}_1|^2}{2 m_0} \ .
\label{EqHamDH}
\eeqn
The first two terms in parentheses represent the 
Kepler Hamiltonian. The third term represents
the motion of the center of mass, which moves as a free 
particle and can be ignored. The last term 
needs to be integrated separately in symplectic 
algorithms with DH coordinates. This term is
often denoted as $H_{\rm Sun}$ because 
$-\v{\P}_1 = \v{p}_0 - (m_0/M)(\v{p}_0 + \v{p}_1)$
is the barycentric momentum of the Sun.

For Jacobi coordinates, one may use a generating function 
$F_2(\v{x},\v{P})$ of the old positions and the new 
momenta:
\beqn
F_2 = \v{P}_0 (m_0 \v{x}_0 + m_1 \v{x}_1)/M
     +\v{P}_1 (\v{x}_1 - \v{x}_0) \ ,
\eeqn
where $(m_0 \v{x}_0 + m_1 \v{x}_1)/M = \v{X}_{S}$ is 
the center of mass.
The relationship between old and new variables then is:
\beq
\q{\prt F_2}{\prt \v{P}_0} & = &  \v{X}_{S} = \v{Q}_0               \\
\q{\prt F_2}{\prt \v{P}_1} & = & (\v{x}_1 - \v{x}_0) = \v{Q}_1      \\
\q{\prt F_2}{\prt \v{x}_0} & = & -\v{P}_1 + \v{P}_0 m_0/M = \v{p}_0 \\
\q{\prt F_2}{\prt \v{x}_1} & = &  \v{P}_1 + \v{P}_0 m_1/M = \v{p}_1
\eeq
The Hamiltonian becomes:
\beqn
H = \left(
     \q{|\v{P}_1|^2}{2\mu}
   - \q{G m_0 m_1}{|\v{Q}_1|} \right)
   + \q{|\v{P}_0|^2}{2M} \ ,
\label{EqHamJ}
\eeqn
where $\mu = m_0 m_1/M$ is the reduced mass. Again,
the first two terms represent the Kepler Hamiltonian
and the third term (center of mass) can be ignored. 
However, comparing the Hamiltonian in Jacobi coordinates
(Eq.~(\ref{EqHamJ})) and in DH coordinates 
(Eq.~(\ref{EqHamDH})) shows that in Jacobi coordinates
the relevant Kepler mass is $\mu$ (rather than $m_1$) and 
$H_{\rm Sun}$ is absent.

\ifTWO
\begin{figure}[tttttt]
\def\epsfsize#1#2{0.45#1}
\hspace*{-.5cm}
\else
\begin{figure}[tttttt]
\def\epsfsize#1#2{0.65#1}
\hspace*{2cm}
\fi
\centerline{\vbox{
\epsfbox{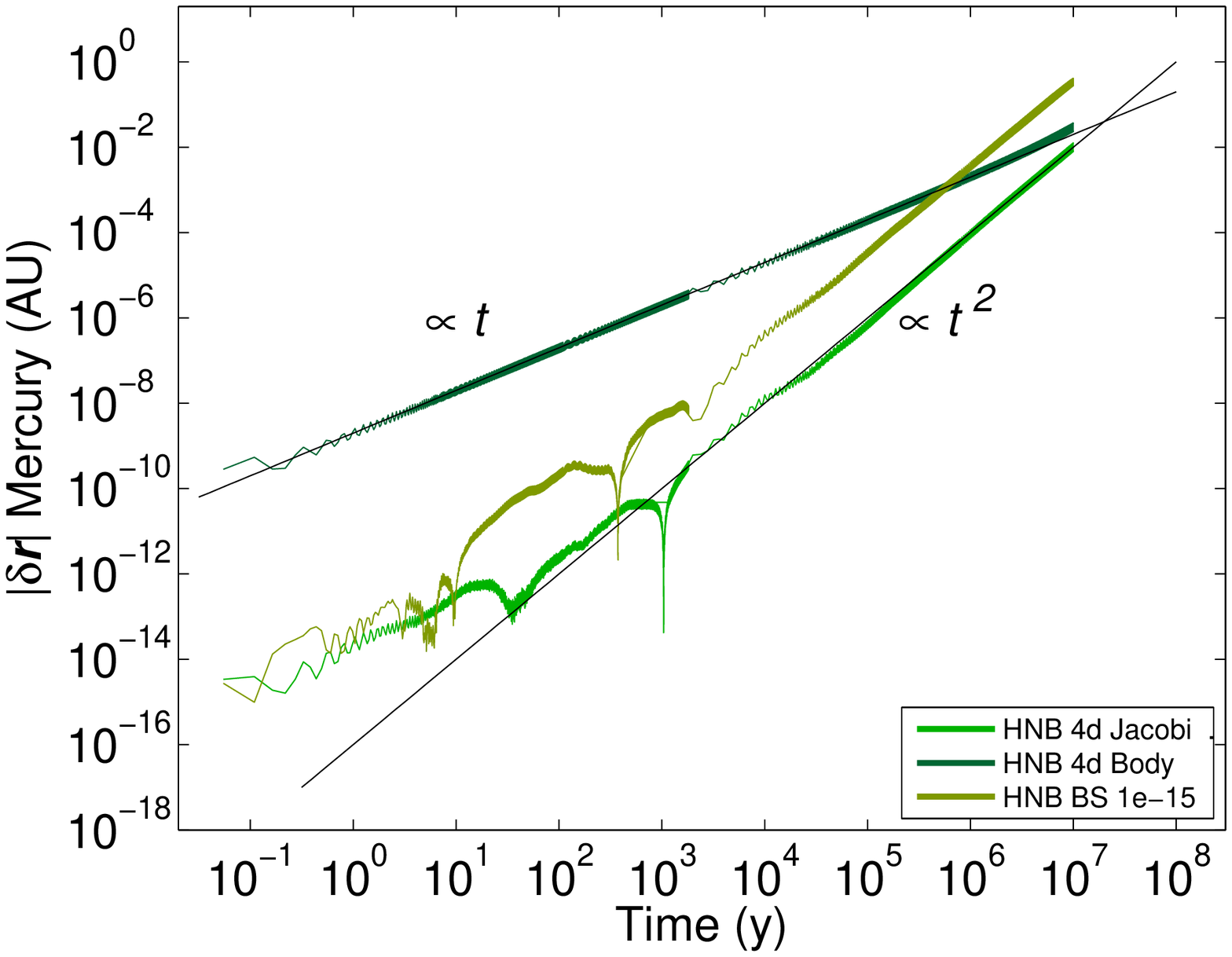}
}}
\caption[]{\scs
Difference between numerically and analytically determined 
position vector of Mercury ($\drM = |\v{r}_{\cal M}^{\rm num}
- \v{r}_{\cal M}^{\rm anl}|$) from 2-body integrations 
with \hnb\ of the Sun and Mercury. Graphs labeled ``Jacobi''
and ``Body'' show \drM\ between the 2nd order symplectic integrator
using Jacobi and bodycentric coordinates (both 4-day timestep)
and the analytical solution. 
BS: \drM\ between non-symplectic Bulirsch-Stoer algorithm 
(relative accuracy set to $10^{-15}$) and analytical solution.
}
\label{FigDR}
\end{figure}

Thus, even for the simple 2-body problem, there is a 
difference between symplectic algorithms with heliocentric 
and Jacobi coordinates. Because the Hamiltonian is purely 
Keplerian in Jacobi coordinates (no solar correction),
symplectic integration of the 2-body problem in Jacobi 
coordinates should be more accurate than in DH coordinates.
Indeed, 10-Myr integrations with \hnb\ of the Sun and
Mercury showed that \drM\ (difference in Mercury's numerical
and analytical position vector) was substantially smaller 
in integrations with Jacobi than DH coordinates 
over the first 1~Myr or so (Fig.~\ref{FigDR}). 
On a timescale of $10^{-1}$~y, the numerical
position error is close to machine precision for Jacobi 
coordinates but $\sim$$10^5$-times larger for DH 
coordinates (both 4-day timestep). Subsequently, 
the position error grows approximately quadratic and 
linear in time for Jacobi and DH coordinates, respectively.
Additional runs with \hnb\ using a non-symplectic
Bulirsch-Stoer algorithm (relative accuracy set to
$10^{-15}$) yields errors in \rM\ closer 
to those of the symplectic integrator with Jacobi 
coordinates (Fig.~\ref{FigDR}). Note, however,
that on timescales of $10^6$--$10^7$~y errors in \rM\
grow to similarly large values in all three integrations.

Given that increases in Mercury's eccentricity are critical 
for the potential destabilization of the Solar System, 
accurate numerical integration of its orbit is key. 
Because Mercury is the innermost planet, one might argue 
that Jacobi coordinates are better suited than DH coordinates 
for this task \citep{duncan98} and that integrating Solar
System orbits (specifically Mercury's orbit) 
using DH coordinates simply lacks the necessary accuracy.
In terms of errors in Mercury's position vector, the 
present 2-body integrations support this notion, 
but only on timescales shorter than $\sim$$10^6$~y.

\subsection{Bulirsch-Stoer algorithm: Eccentricity errors
\label{SecBS}}

As described above, I also tested whether the statistically 
different \eMx\ in the 500-Myr runs (high initial \eM)
are related to typically larger
errors associated with bodycentric vs.\ Jacobi coordinates
at the same step size. I repeated the 500-Myr runs 
with the bodycentric \hnb\ setup but an eight-fold smaller 
timestep (0.25~d, $\sim$100-fold smaller $|\Delta E/E|$).
The basic result remained the same. At high
initial \eM, the setup using bodycentric coordinates leads 
to significantly smaller mean \eMx\ than the setup using Jacobi
coordinates (Fig.~\ref{FigEmax}). This result suggests that 
the effect of different integration coordinates on the 
statistics of Mercury's eccentricity evolution are not 
caused by errors arising from a too large timestep in the 
bodycentric setup.

Further tests can be performed by comparing the results of the
symplectic integrators to results of non-symplectic 
integrators, e.g.\ the Bulirsch-Stoer (BS) algorithm. Note
that such tests are usually run over shorter time 
intervals as in most cases non-symplectic integrations 
are computationally much more expensive than 
symplectic integrations. Comparison between \hnb's
BS and the 2nd order symplectic integration of the 
eight planets and Pluto starting at initial $\eM = 0.53$
showed a moderate rise in \DeM\ over 10~Myr when using
Jacobi coordinates and a 2-day timestep
($\max|\DeM| \ \lesssim 10^{-4}$, Fig.~\ref{FigBS}a). 
This was not the case for bodycentric coordinates/2-day 
timestep vs.\ BS, where $\max|\DeM|$
grew rapidly to $\sim$$10^{-2}$ over 10~Myr (Fig.~\ref{FigBS}b). 
However, a moderate rise was found again in \DeM\ over 
10~Myr when using bodycentric coordinates and a 0.25-day 
timestep ($\max|\DeM| \ \lesssim 10^{-4}$, Fig.~\ref{FigBS}c).
Thus, in seeming contradiction to the results of the 500-Myr 
runs (see above), the comparison to BS suggests that 
differences in \eM\ may be rectified provided that the 
timestep in the bodycentric setup is ``sufficiently
small''. If so, then what is sufficiently small?

\ifTWO
\begin{figure}[tttttt]
\def\hsp{-1cm}
\def\epsfsize#1#2{0.50#1}
\else
\begin{figure}[tttttt]
\def\hsp{4cm}
\def\epsfsize#1#2{0.45#1}
\fi
\centerline{\vbox{
\hspace*{\hsp}
\epsfbox{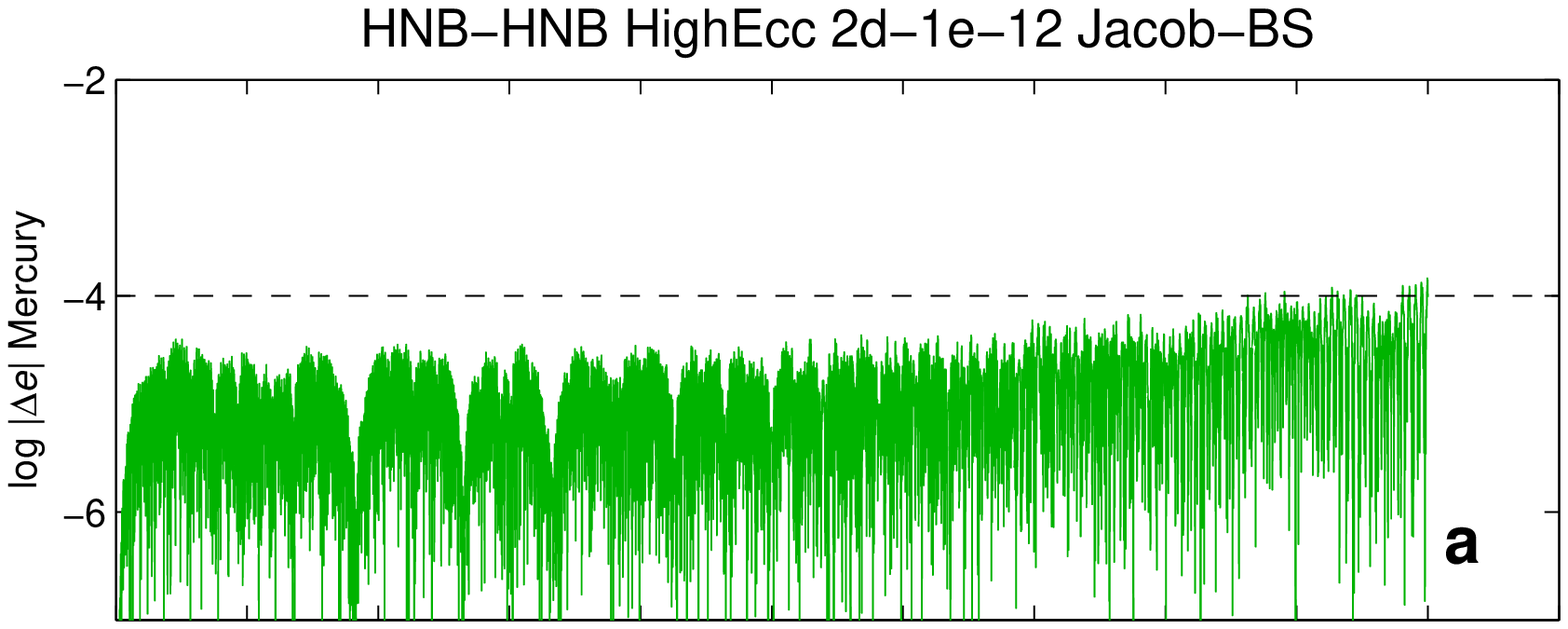}}}     
\centerline{\vbox{
\hspace*{\hsp}
\epsfbox{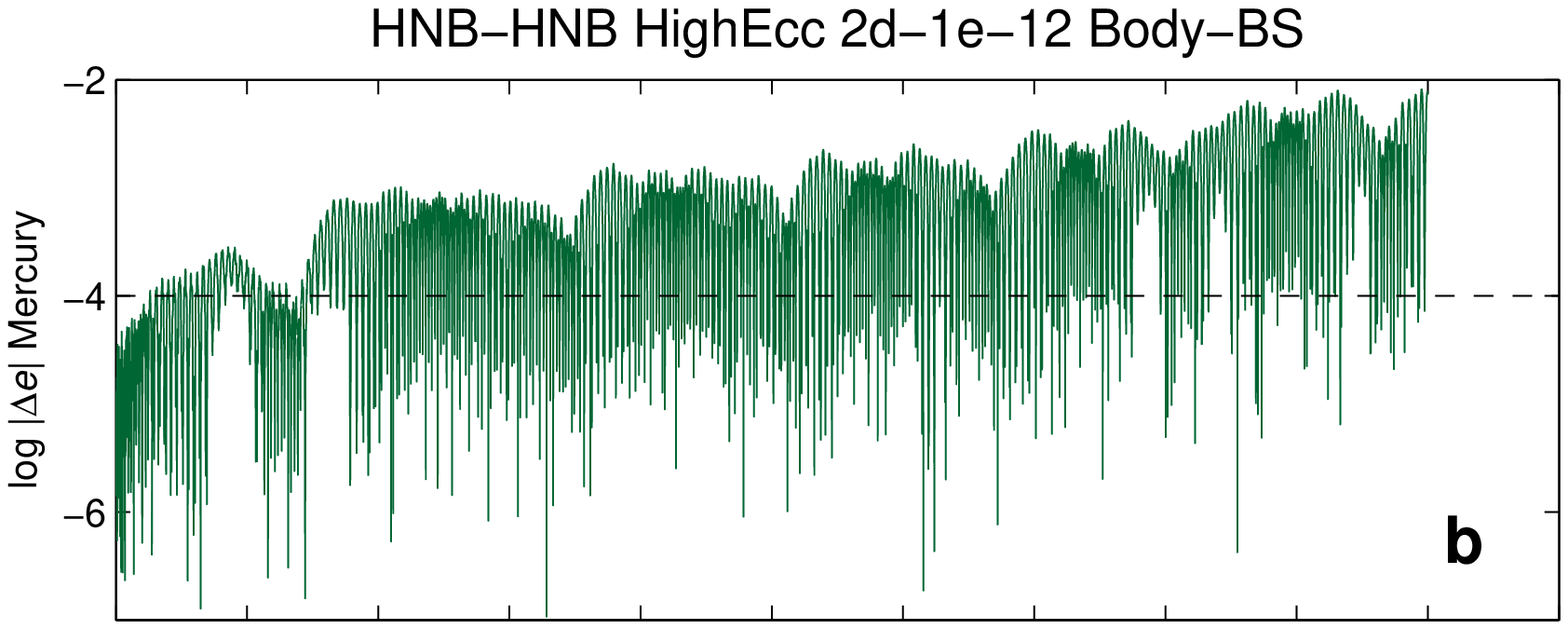}}}     
\centerline{\vbox{
\hspace*{\hsp}
\epsfbox{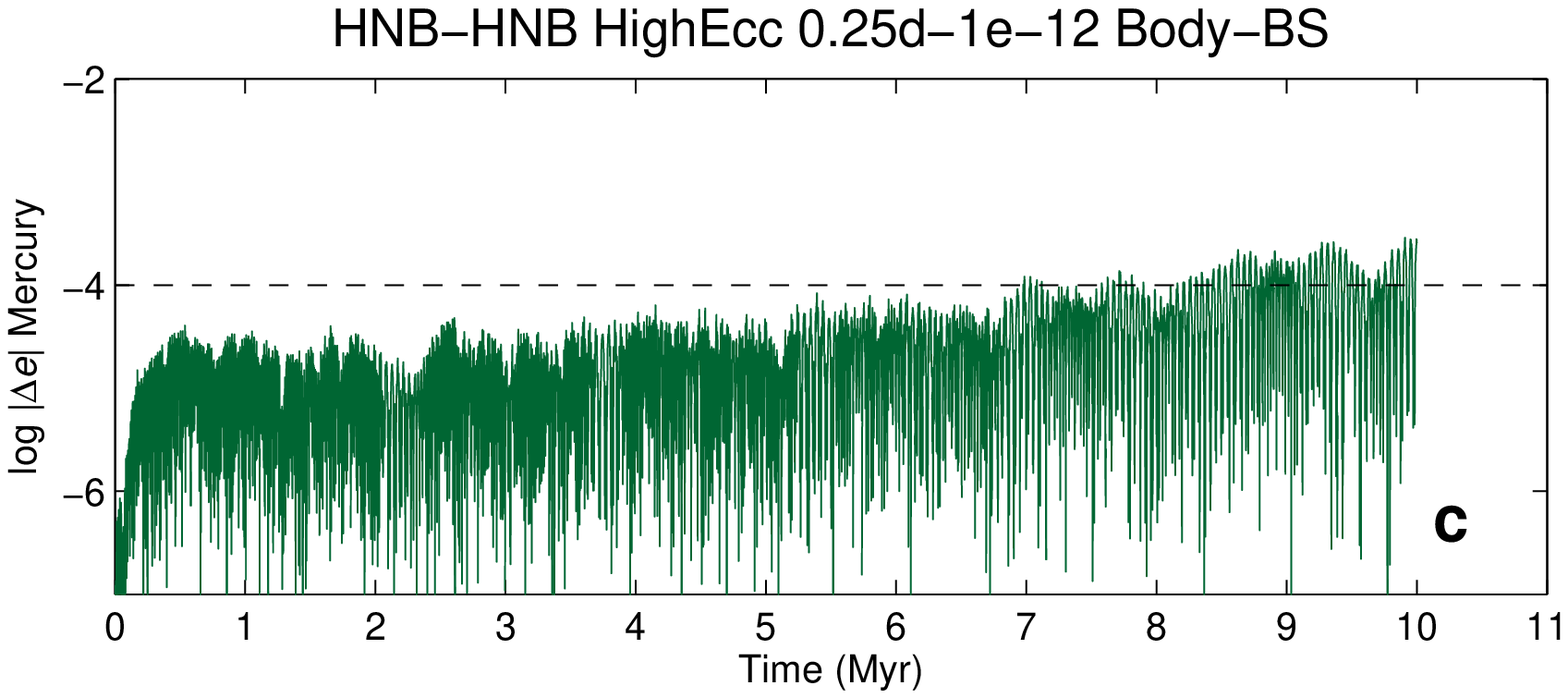}}}    
\caption[]{\scs
Computed difference in Mercury's eccentricity (\DeM) between
two runs each per panel for different integrations and options.
All runs with \hnb\ (HNB) including the eight planets and Pluto.
Body: bodycentric, Jacob: Jacobi coordinates, BS: Bulirsch-Stoer. 
HighEcc: initial 
$\eM = 0.53$ (see text). (a) 2nd order symplectic algorithm 
with 2-day timestep and Jacobi coordinates vs.\ non-symplectic 
Bulirsch-Stoer algorithm (relative accuracy set to $10^{-12}$).
(b) Same as (a) but bodycentric coordinates. (c) Same as (b) but 
0.25-day timestep.
}
\label{FigBS}
\end{figure}

For the symplectic 2-body integrations discussed above, a 
timestep smaller than $\sim$$0.05$~days would achieve 
roughly the same accuracy with bodycentric coordinates as 
with Jacobi coordinates and a 4-day timestep. If the same 
timestep ratio were to apply to full Solar System integrations,  
then the timestep would have to be reduced by a factor of 
80 when using bodycentric coordinates instead of Jacobi 
coordinates. For example, the 500-Myr runs with 2-day
timestep ran over roughly 2.2~days wall-clock 
time. Thus, repeating those runs with an 80-times reduced
timestep would take about 180~days (2.6 years for the 
5-Gyr runs). Not only are such runs currently impractical,
at small timesteps one also needs to consider 
accumulation of numerical errors, which typically scale with
the number of steps. Finally, could the accuracy of
symplectic integrations be tested by comparison to 
long-term integrations using non-symplectic 
algorithms such as Bulirsch-Stoer? As mentioned above, 
the Bulirsch-Stoer algorithm is currently computationally 
too expensive for long-term integrations. In addition, 
non-symplectic algorithms usually suffer from substantial 
long-term drifts in total energy and angular momentum.

\subsection{Sun + 5 planets: Statistics at high \eM
\label{SecSp5}}

While long-term integrations of the full equations of 
motion of the complete Solar System
at very small timesteps appear impractical at 
this stage, insight into algorithm performance may be 
gained from test integrations of planetary 
systems of somewhat reduced complexity.
It turned out that the $(g_1$$-$$g_5)$ resonance pattern
at high initial \eM\ can be reproduced with just a 6-body
setup 
\ifTWO
\newpage
\vspace*{10cm}
\noindent
\fi
(Sun + 5 planets: Mercury, Venus, Earth, Jupiter, 
and Saturn). When contributions from general relativity 
are ignored, this setup frequently leads to rapid \eM\ increases
after $\sim$10~Myr. The estimated Lyapunov time for
Mercury's orbit in this system is $\sim$0.6~Myr. 
Mercury's orbit is unstable and intermittently 
switches between resonant and non-resonant phases 
(Fig.~\ref{FigEI12myr}). 
The resonant phase is 
associated with the $(g_1$$-$$g_5)$ resonance and 
typically high values in Mercury's inclination
(ca.\ $10^\circ$--$20^\circ$). The 
non-resonant phase is typically associated with a drop in 
Mercury's inclination and eccentricity. After about
$8$--$10$~Myr, \DeM\ between two nearby orbits reaches
the magnitude of \eM\ itself, which subsequently increases
beyond 0.8 between $10$--$12$~Myr in some simulation but not 
in others, depending (sensitively) on initial conditions.
Hence, the system may provide some useful algorithm 
tests and allow statistical analyses over only a 12-Myr 
interval --- an interval short enough for integrations 
with a very small timestep.

\ifTWO
\begin{figure}[t]
\def\epsfsize#1#2{0.70#1}
\hspace*{-8cm}
\else
\begin{figure}[tttttt]
\def\epsfsize#1#2{0.70#1}
\hspace*{1cm}
\fi
\centerline{\vbox{\epsfbox{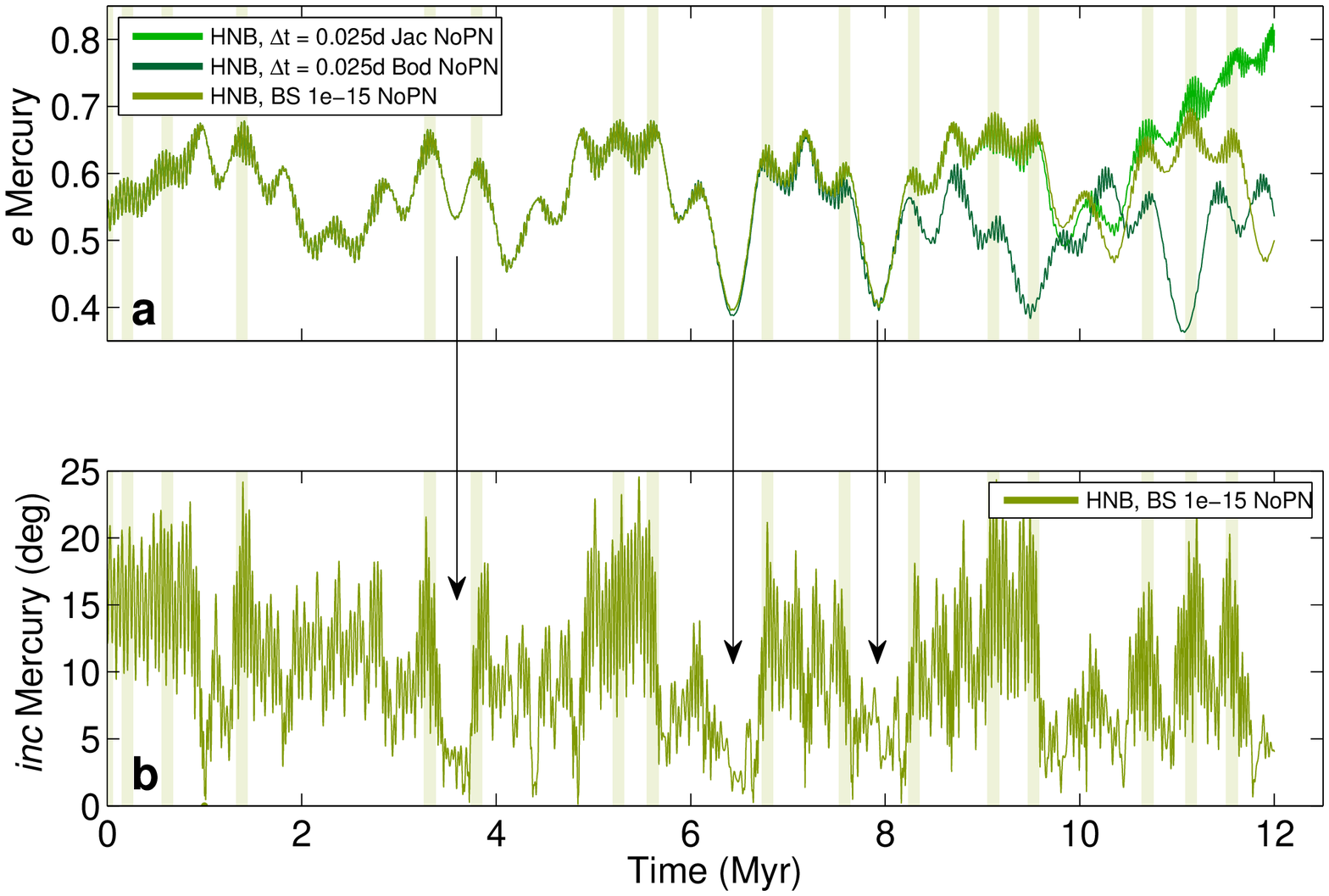}}}
\caption[]{\scs Mercury's eccentricity and inclination
from 12-Myr runs (Sun + 5 planets) at high initial \eM\ without 
contributions from general relativity (NoPN). Jac: Jacobi, 
Bod: bodycentric, coordinates, $\Dt$: step size in days, 
BS: Bulirsch-Stoer (relative accuracy set to $10^{-15}$).
(a) One example from each of three setups (see legend,
total \# of runs for each setup = 40). (b) Mercury's inclination 
for the BS solution corresponding to (a). Vertical bars
highlight several eccentricity maxima and high inclination
values associated with the $(g_1$$-$$g_5)$ resonance
(BS solution). Arrows indicate a few examples of eccentricity 
minima and low inclination values associated with non-resonant 
phases.
}
\label{FigEI12myr}
\end{figure}

\def\mm{0.683}
\ifTWO
\begin{figure}[tttttt]
\def\epsfsize#1#2{0.25#1}
\def\hsp{-.5cm}
\else
\begin{figure}[tttttt]
\def\epsfsize#1#2{0.40#1}
\def\hsp{1cm}
\fi
\centerline{
\epsfbox{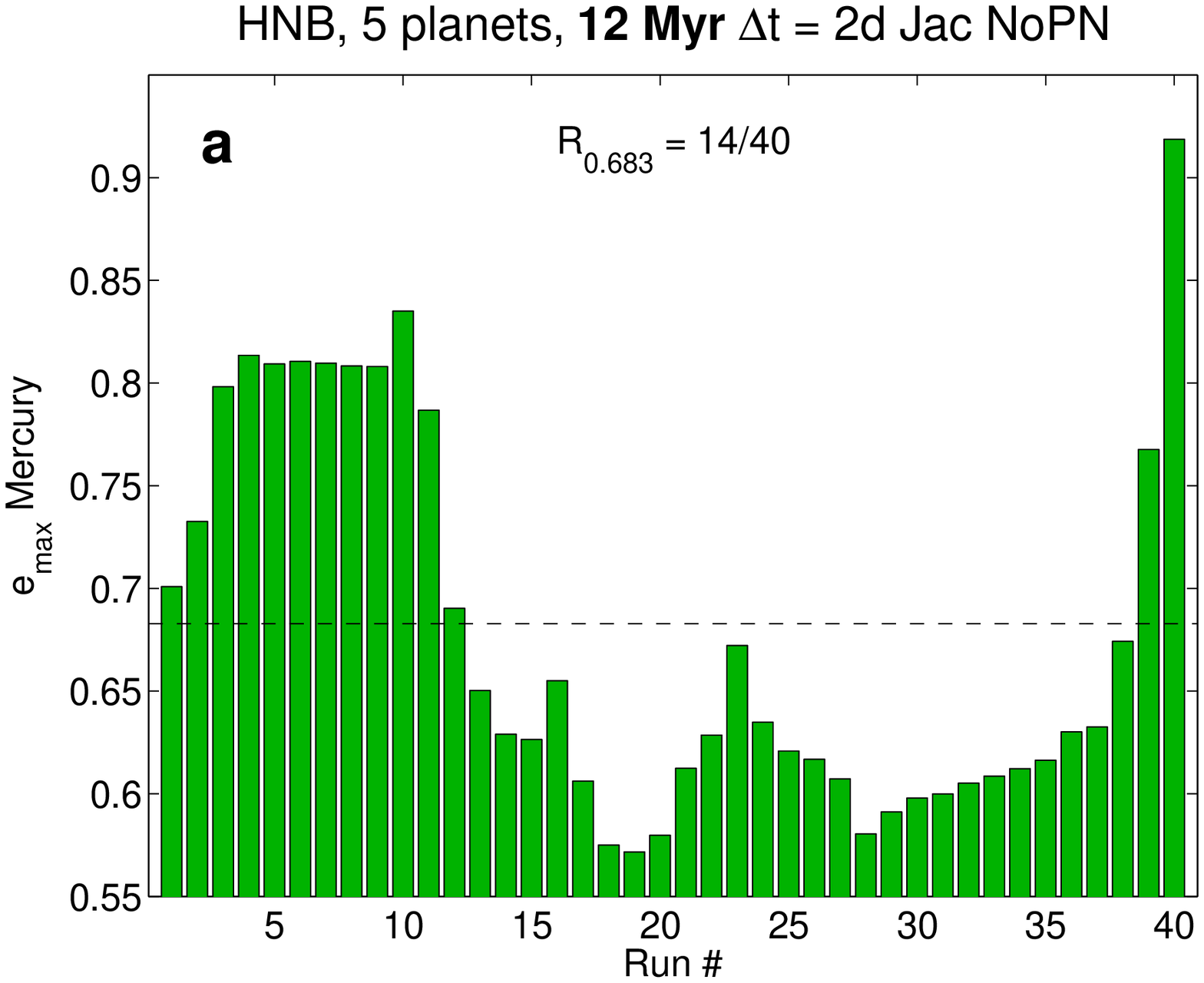}
\epsfbox{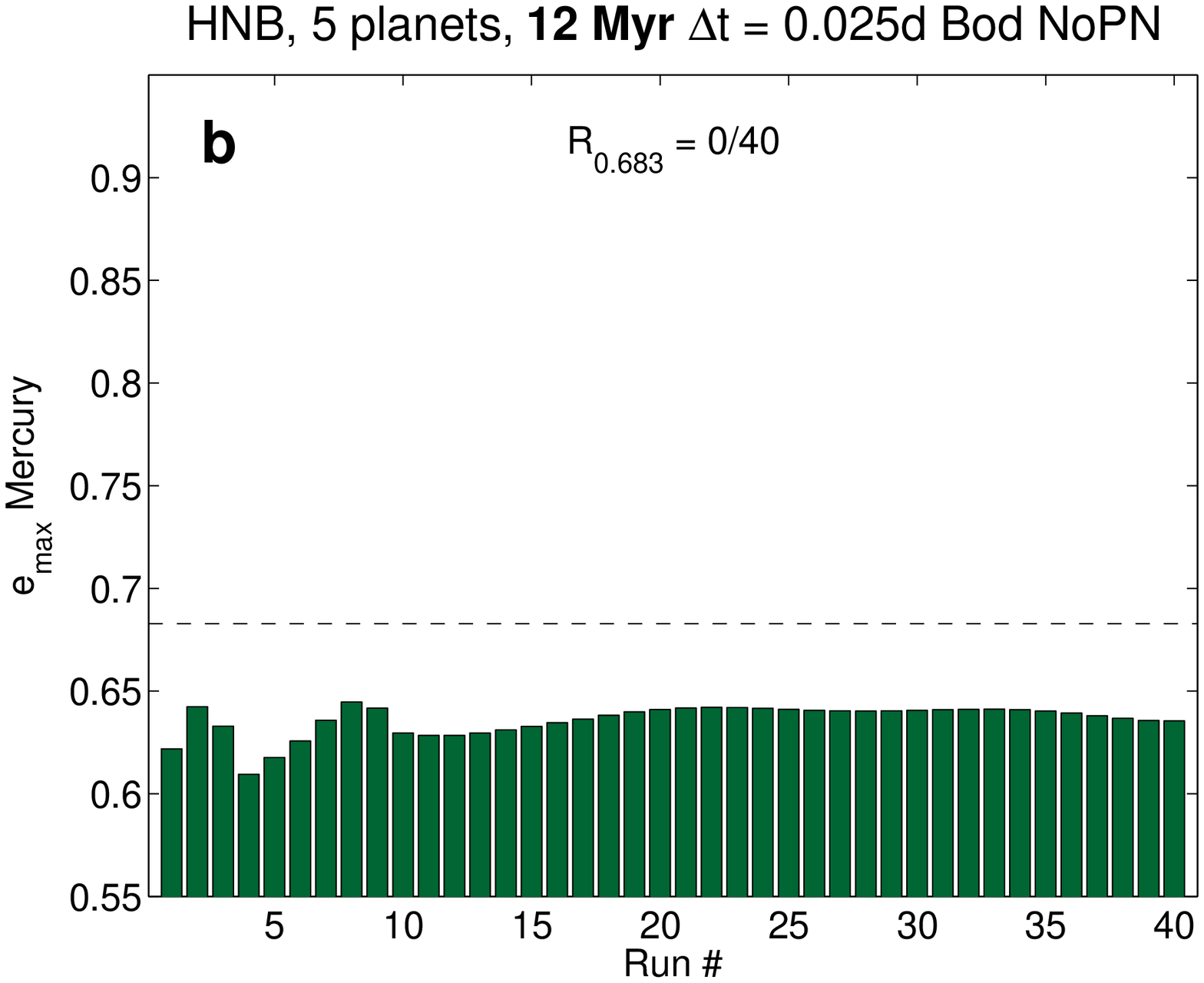}
}
\centerline{
\epsfbox{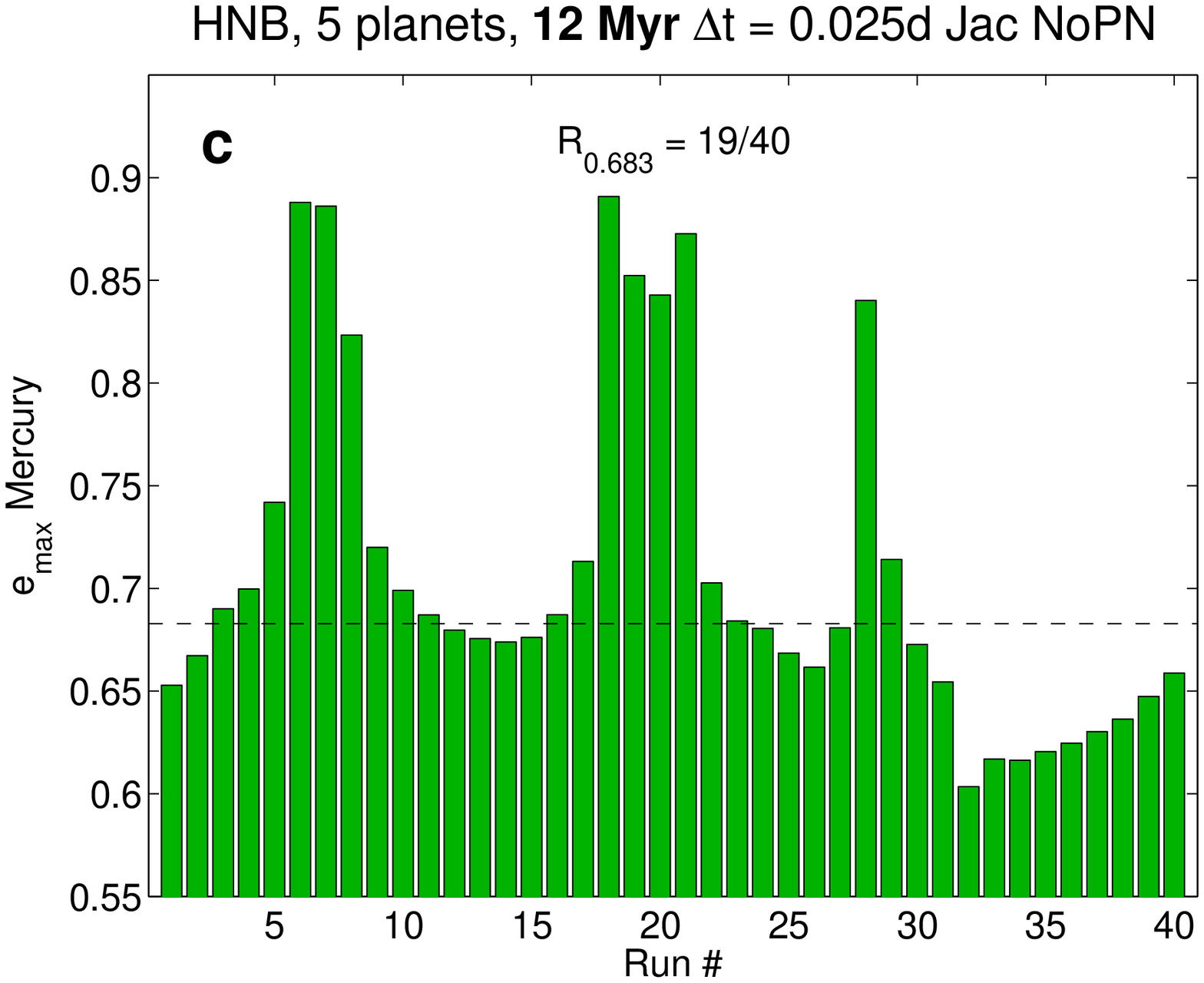}
\epsfbox{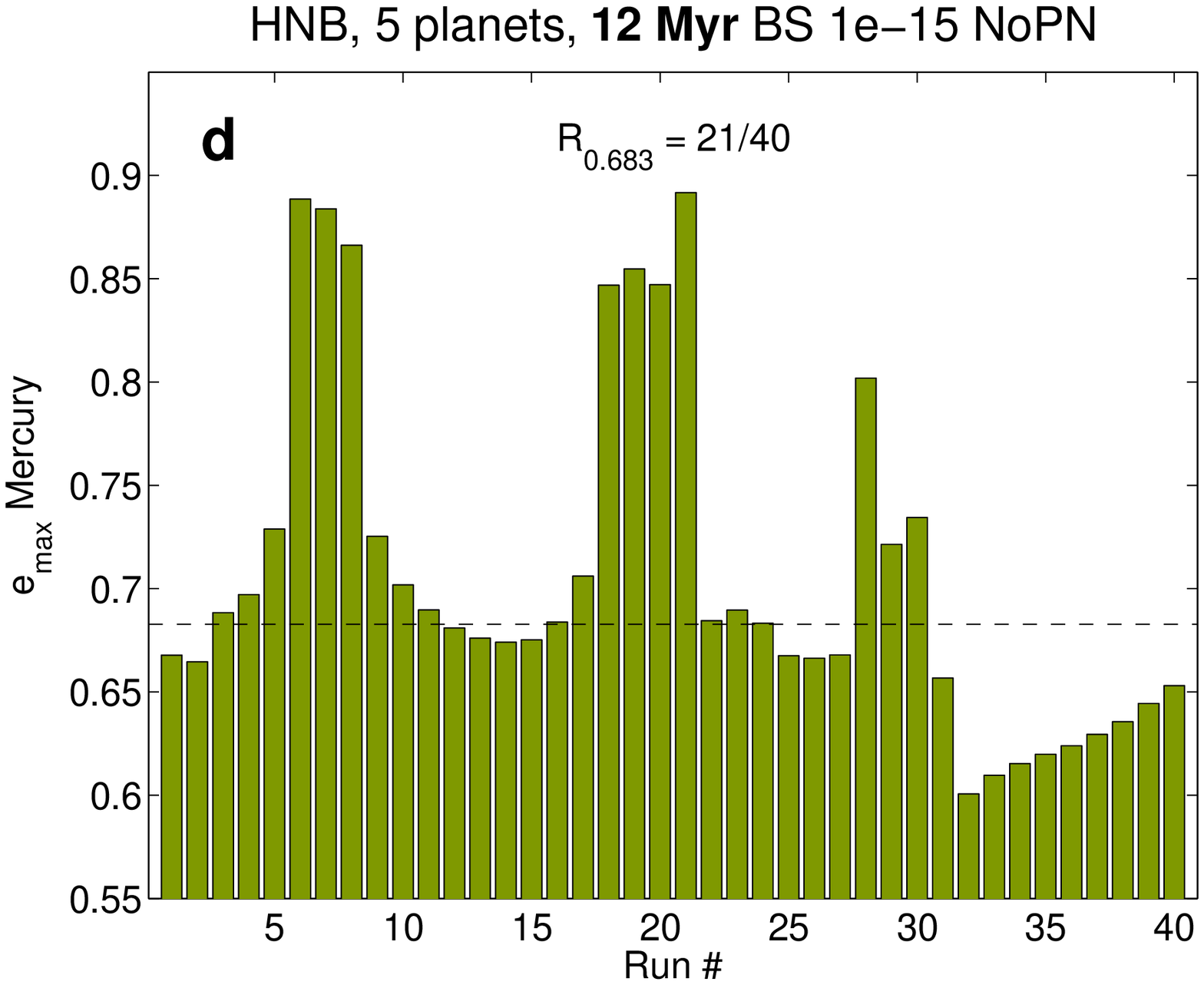}
}
\caption[]{\scs Mercury's maximum eccentricity (\eMx) 
achieved during final 2~Myr of 12-Myr runs (Sun + 
5 planets, see text). 
Jac: Jacobi, Bod: bodycentric, coordinates, $\Dt$:
step size in days,
BS: Bulirsch-Stoer (relative accuracy set to $10^{-15}$),
NoPN: No contributions from general relativity.
$R_{\mm}$ indicates the number of solutions in 
which \eM\ grows beyond \mm\ (mean of all 160 runs,
dashed lines).}
\label{FigEmaxR36}
\end{figure}

I have integrated 40 solutions each with four different 
numerical setups over 12 Myr using \hnb\ (Fig.~\ref{FigEmaxR36}): 
(a) symplectic algorithm with 2-day timestep 
and Jacobi coordinates,
(b) symplectic algorithm with 0.025-day timestep 
and bodycentric coordinates,
(c) symplectic algorithm with 0.025-day timestep 
and Jacobi coordinates,
(d) Bulirsch-Stoer (BS) algorithm (relative accuracy set to 
$10^{-15}$). All simulations ignore contributions 
from general relativity; the symplectic integrators are 
2nd order schemes. The setups (b) and (c) used an exact 
timestep of $0.025390625$~d (finite binary representation
to minimize round-off errors),
which is about 80 times smaller than the 2-day timestep of
setup (a) (see discussion Section~\ref{SecBS}).
The four sets of integrations ($N = 40$ each) started 
from the same set of initial conditions, where Mercury's 
initial radial distance was offset by 10~m between every 
two adjacent orbits (largest overall offset: 
$\rm39\x10~m=390$~m).

For all simulations, Mercury's maximum eccentricity (\eMx) 
achieved during the final 2~Myr of the 12-Myr runs was
recorded, providing a metric for either an increase 
or a decline in \eMx\ towards the end of the integration 
(Fig.~\ref{FigEI12myr}). Of the four numerical 
setups, only the symplectic integrator with 
$\Dt = 0.025$~d/Jacobi coordinates and the Bulirsch-Stoer 
algorithm yield similar statistical results for \eMx\
(Fig.~\ref{FigEmaxR36}). In comparison, the symplectic 
integrator with $\Dt = 2$~d/Jacobi coordinates shows 
a reduced tendency for large \eM\ increases and has
a lower mean \eMx\ value. Strikingly, the symplectic integrator 
with bodycentric coordinates predicts a decline in \eMx\
to values below 0.65 between $10$--$12$~Myr in all simulations 
(0.65 is roughly equal to the \eMx\ value from $0$--$10$~Myr,
see Fig.~\ref{FigEI12myr}).

For this particular system, the statistical results
shown in Fig.~\ref{FigEI12myr} have important implications.
Based on the agreement between BS and the symplectic 
integrations with $\Dt = 0.025$~d/Jacobi coordinates, it 
is likely that these two setups provide the most 
accurate results. The first implication then is that 
$\Dt = 2$~d is too large for a 2nd order symplectic integrator 
with Jacobi coordinates at high \eM.
For instance, the calculated odds for \eMx\ to increase beyond 
0.683 (mean of all 160 runs) are 48\% at $\Dt = 0.025$~d but
only 35\% at $\Dt = 2$~d. Second, the symplectic 
integrator with bodycentric coordinates yields incorrect
results for this system even at a very small time step 
of $\rm 0.025~d = 36$~min!
The tendency to underestimate increases in Mercury's
eccentricity at high \eM, which
was also observed in the 500-Myr integrations of the
full Solar System (Fig.~\ref{FigDEcc}), is therefore unlikely 
related to errors associated with the size of the 
timestep. Fundamental differences in the implementation of 
different integration coordinates in symplectic schemes are
more likely to be the cause (Section~\ref{SecTwoBdy}).

The results of the 12-Myr runs also help resolving the 
apparent contradiction between the 500-Myr 
statistics and the \eM\ comparison to BS mentioned in 
Section~\ref{SecBS}. On the one hand, the 500-Myr results 
suggest that the different \eMx\ statistics are largely
independent of the step size in the bodycentric
setup (Fig.~\ref{FigEmax}). On the other hand, the 
comparison to BS suggests that differences in \eM\ can 
be minimized when the step size in the bodycentric 
setup is reduced (Fig.~\ref{FigBS}). In fact, for 
the 12-Myr runs both statements are correct. At a 
small timestep of $0.025$~d, \DeM\ (BS minus Jacobi 
setup) and \DeM\ (BS minus bodycentric setup) are 
virtually identical during the first 4~Myr (not shown). 
In this interval, \DeM\ is dominated by 
polynomial growth. During the final 2~Myr, however, 
all runs of the bodycentric setup show a decline in 
\eMx\ (contrary to the BS and Jacobi setup).
In this interval, \DeM\ is dominated 
by exponential growth. Importantly, the final interval 
determines the statistics of Mercury's ultimate eccentricity 
evolution. Thus, a smaller step size does improve the 
bodycentric setup's accuracy during polynomial growth of 
\DeM\ but has little effect during exponential growth 
of \DeM. 

Note that the implications outlined above strictly only 
apply to the system of reduced complexity studied in this 
section. That is, a system comprising the Sun and just five 
planets with specific initial conditions (including high 
initial \eM), integrated over only 12~Myr and ignoring 
Post-Newtonian corrections.

\section{Conclusions}

Reliable predictions of the Solar System's dynamic stability 
over billions of years not only require integrators that
are fast and accurate but also produce robust statistical 
results in ensemble integrations. Ensemble integrations
are necessary because of the chaotic behavior
of the system. Currently, symplectic integrators are 
probably the best, if not the only, choice in terms of 
speed and accuracy. However, the present results show that 
tackling statistics is trickier as, for instance, the 
predicted probability for a large increase in Mercury's 
eccentricity (\eM) depends on the choice of integration 
coordinates in symplectic algorithms. 
Several tests performed here suggest that 
using Jacobi coordinates in symplectic integrations of the Solar 
System is more reliable than using bodycentric coordinates
when Mercury's orbital eccentricity is high. However, 
these tests reveal little about the statistics of 
long-term integrations
when Mercury's initial orbital eccentricity is low.
In fact, the influence of Jacobi and bodycentric coordinates 
on the statistics of Mercury's eccentricity evolution 
appears to be opposite over 5~Gyr at low initial \eM\
than over 500~Myr at high initial \eM. Moreover, even
accepting superiority of Jacobi coordinates at high 
\eM, what is the proper step size to obtain robust 
statistical results? If applicable, the results of the 
system of reduced complexity (Section~\ref{SecSp5}) suggest 
a timestep that could be as small as 0.025~d. Even
when starting with a larger timestep and
reducing the timestep during the symplectic integration
(which should be avoided), such small timesteps would
still pose a challenge in terms of integration time.
Then do symplectic integrations with Jacobi coordinates
and typical timesteps of several days underestimate the
odds for disaster (Fig.~\ref{FigEmaxR36})? Again,
if applicable, such simulations would underestimate 
the odds for disaster once \eM\ has already reached 
values $>$0.5 or so. But it is not clear whether
the odds are predicted correctly to reach those values
in the first place when starting at low \eM.

It appears that several centuries 
of analytical research, modern state-of-the-art numerical 
algorithms, and current CPU power has brought us 
closer to answering the question of the Solar System's 
long-term stability. However, a definite, robust answer 
still seems to be lacking, including answers based on 
statistical approaches. It is likely that the odds for
the catastrophic destabilization of the inner planets 
are in the order of a few percent. But what percentage 
exactly?


\vfill
\noindent
{\small
{\bf Acknowledgments.} 
I thank the anonymous reviewer for constructive comments and 
Don David Ho and B.~R.~Oppenheimer for helpful conversations.
I am grateful to Peter H. Richter who dared to introduce
us to Chaos, Poincar{\'e}, and solar system dynamics
in a 1989-undergraduate physics course on classical 
mechanics.
}

\appendix

\section{Contributions from general relativity}

Relativistic corrections are critical as they substantially 
reduce the probability for Mercury's orbit to achieve large
eccentricities \citep{laskar09}. General relativity (GR)
corrections are available 
in \hnb\ but not in \ms. Post-Newtonian (PN) corrections for 
symplectic integration \citep{mikkola98,soffel89} were 
therefore implemented before using \ms. For more information 
on symplectic algorithms, see \citet{wisdom91,saha92,chambers99}.
Note that the 
PN code was fully embedded in the symplectic hybrid integrator 
\verb|mdt_hy()|, and not just added as an auxiliary force 
term in \verb|mfo_user()|. For every planet, one needs to 
solve an equation of the form \citep{mikkola98}:
\beqn
\v{a} = \ddot{\v{r}} = \v{F}
                   + \v{f}_0(\v{r},t) 
                   + \q{1}{c^2} \ \v{f}_1(\v{r},\v{v}) \ ,
\eeqn
where $\v{F}$ is the two-body acceleration, 
$\v{f}_0(\v{r},t)$ is the Hamiltonian perturbation, 
$c$ is the speed of light, and $\v{f}_1(\v{r},\v{v})$ is 
the expression for the first Post-Newtonian contribution 
\citep{soffel89}. Combining the implicit midpoint method 
and generalized leapfrog, and denoting the timestep as $h$,
one obtains for the velocity jumps \citep{mikkola98}:
\beqn
\dv =   h \v{f}_0(\v{r},t) 
      + \q{h}{c^2} \ 
        \v{f}_1 \left( \v{r},\v{v}+ \qt\dv \right) \ ,
\label{Eqdv}
\eeqn
which must be solved for \dv.
The first PN acceleration term may be written as 
\citep{soffel89}:
\beqn
\q{1}{c^2} \v{f}_1(\v{r},\v{v}) = 
          \q{G M}{c^2} \left[
      -   \q{\v{v}^2}{r^3} \ \v{r}
      + 4 \q{G M}{r^4}     \ \v{r}
      + 4 \q{(\v{r} \cdot \v{v})}{r^3} \ \v{v}
      \right] \ .
\eeqn
Inserting into Eq.~(\ref{Eqdv}) yields:
\beqn
\dv & = &   h \ \v{f}_0(\v{r},t) \\
    &   & + h \ 
                 \q{G M}{c^2} \left[
             -   \q{(\v{v}+\qt\dv)^2}{r^3} \ \v{r}
             + 4 \q{G M}{r^4}     \ \v{r}
             + 4 \q{[\v{r} \cdot (\v{v}+\qt\dv)]}{r^3} \ (\v{v}+\qt\dv)
              \right] \ ,
\eeqn
which was solved iteratively for \dv. Because the relativistic 
term is small, convergence is rapid (usually 2 iterations),
and $[\dv(i=2)-\dv_0]/\dv_0 < 5\e{-16}$ for Mercury.
The computational overhead for including GR contributions 
in \ms\ as described above was about 15-20\% (wall-clock 
time). Over the 21st century, Mercury's average perihelion 
precession (only due to GR) was 0.42976''~y\pmo\ computed with
\hnb\ and 0.42978''~y\pmo\ computed with \ms\ and the above
GR implementation. In terms of Mercury's long-term eccentricity 
(\eM) evolution, the difference in \eM\
between \ms\ and \hnb\ runs (both with GR correction, 
$\D t = 6$~days, and bodycentric coordinates) was 
$\sim$$10^{-4}$ over the first 40~Myr.

%
%

\renewcommand{\baselinestretch}{1.2}

\begin{table}[hhhhhh]
\begin{center}
\caption{Initial conditions of the eight planets and Pluto$^a$ 
for 5-Gyr runs from DE431. Heliocentric positions \v{r} (AU) and 
velocities \v{v} (AU~d\pmo). \label{TabDE431}}
\begin{tabular}{lccc}
\tableline\tableline
 & $x$ & $y$ & $z$ \\
\tableline
 & & Mercury & \\
\v{r} &
 $-$1.40712354144735680E-01 & $-$4.43906230277241465E-01 & $-$2.33474338281349329E-02 \\
\v{v} &
  +2.11691765462179472E-02 & $-$7.09701275933066148E-03 & $-$2.52278032052283448E-03 \\
 & & Venus & \\
\v{r} &
 $-$7.18629835259113170E-01 & $-$2.25188858612526514E-02 & +4.11716131772919824E-02  \\
\v{v} &
  +5.13955712094533914E-04 & $-$2.03061283748202266E-02 & $-$3.07198741951420558E-04 \\
 & & Earth + Moon & \\
\v{r} &
 $-$1.68563248623229384E-01 &  +9.68761420122898564E-01 & $-$1.15183154209270563E-06 \\
\v{v} &
 $-$1.72299715055074729E-02 & $-$3.01349780674632205E-03 & +2.41254068070491868E-08 \\
 & & Mars & \\
\v{r} &
  +1.39036162161402177E+00 & $-$2.09984400533893799E-02 & $-$3.46177919349353047E-02 \\
\v{v} &
  +7.47813544105227729E-04 & +1.51863004086334515E-02 &  +2.99756038504512547E-04 \\
 & & Jupiter & \\
\v{r} &
  +4.00345668418424960E+00 &  +2.93535844833712467E+00 & $-$1.01823217020834328E-01 \\
\v{v} &
 $-$4.56348056882991196E-03 & +6.44675255807273997E-03 &  +7.54565159392195741E-05 \\
 & & Saturn & \\
\v{r} &
  +6.40855153734800886E+00 &  +6.56804703677062207E+00 & $-$3.69127809402511886E-01 \\
\v{v} &
 $-$4.29112154163879215E-03 &  +3.89157880254167561E-03 &  +1.02876894772680478E-04 \\
 & & Uranus & \\
\v{r} &
  +1.44305195077618524E+01 & $-$1.37356563056406209E+01 & $-$2.38128487167790809E-01 \\
\v{v} &
  +2.67837949019966498E-03 &  +2.67244291355153403E-03 & $-$2.47764637737944378E-05 \\
 & & Neptune & \\
\v{r} &
  +1.68107582839480649E+01 & $-$2.49926499733276124E+01 &  +1.27271208982211476E-01 \\
\v{v} &
  +2.57936917068014599E-03 &  +1.77676956230748452E-03 & $-$9.59089132565213410E-05 \\
 & & Pluto & \\
\v{r} &
 $-$9.87686582399026491E+00 & $-$2.79580297772433077E+01 &  +5.85080284687055574E+00 \\
\v{v} &
  +3.02870206449818878E-03 & $-$1.53793257901232473E-03 & $-$7.12171623386267461E-04 \\
\tableline
\end{tabular}
\tablenotetext{a}{Masses (Mercury to Pluto in solar masses):
1.66013679527193035E-07,
2.44783833966454472E-06,
3.04043264626852573E-06,
3.22715144505387430E-07,
9.54791938424322164E-04,
2.85885980666102893E-04,
4.36625166899970042E-05,
5.15138902053549668E-05,
7.40740740740740710E-09.
}
\end{center}
\end{table}

\end{document}